\newcounter{parentequation}
\def\beglet{
  \addtocounter{equation}{1}%
  \setcounter{parentequation}{\value{equation}}%
  \setcounter{equation}{0}%
  \def\theequation{\arabic{parentequation}\alph{equation}}%
  \ignorespaces
}
\def\endlet{
  \setcounter{equation}{\value{parentequation}}%
  \def\theequation{\arabic{equation}}%
}
\def\spose#1{\hbox to 0pt{#1\hss}}
\def\approxlt{\mathrel{\spose{\lower 3pt\hbox{$\sim$}}
        \raise 2.0pt\hbox{$<$}}}
\def\approxgt{\mathrel{\spose{\lower 3pt\hbox{$\sim$}}
        \raise 2.0pt\hbox{$>$}}}

\def\today{\ifcase\month\or January\or February\or March\or April\or May\or
      June\or July\or August\or September\or October\or November\or December\fi
      \space\number\day, \number\year}

\def\kms{{\rm km s}^{-1}}

\def\keV{{\rm\thinspace keV\ }}   
     
\def\kg{{\rm\thinspace kg}}     
\def\kpc{{\rm\thinspace kpc\ }}           
     
\def\m{{\rm\thinspace m}}       
\def\Mpc{{\rm\thinspace Mpc\ }}   
\def\Msun{\hbox{$\rm\thinspace M_{\odot}$}}

\def\LCDM{$\Lambda\mbox{CDM}$  }

\def\LWDM{$\Lambda\mbox{WDM}$  }
\def\beq{\vspace{2mm} \begin{equation}}
\def\eeq{\vspace{2mm} \end{equation}}
\def\beqn{\vspace{2mm} \begin{eqnarray}}
\def\eeqn{\vspace{2mm} \end{eqnarray}}

\documentclass[11pt]{mn2e}        

\usepackage{epsfig,multirow}
\usepackage{myaasmacros}
\usepackage{astrobib2}
\begin{document}        
\hsize=6truein  

\title[Elliptical Galaxy Formation]   
{Cosmological Simulations of Elliptical Galaxy Formation in \LCDM and
\LWDM Cosmologies.}         
\author[Lisa J. Wright, Jeremiah P. Ostriker, Thorsten Naab \& George Efstathiou]        
{\parbox[]{7in}
{Lisa J. Wright$^1$, Jeremiah P. Ostriker$^{1,2}$, Thorsten Naab$^{1}$
\& George Efstathiou$^1$ }\\
\\
1. \it Institute of Astronomy, Madingley Road, Cambridge CB3 0HA, UK\\
2. Department of Astrophysics, Peyton Hall, Princeton, USA}

\date{Accepted ???. Received ???; in original form \today}

\maketitle

\begin{abstract}
We present the results of a series of gas dynamical cosmological
simulations of the formation of individual massive field galaxies in
the standard concordance \LCDM and in a \LWDM cosmology each with
$\Omega_0$=0.3 and $\Lambda_0$=0.7. Two high resolution simulations
($2\times 50^3$ gas and dark matter particles) have been performed and
investigated in detail. The gas component was represented by Smooth
Particle Hydrodynamics (SPH) and a simple star formation algorithm was
applied. The galaxies form in an initial burst of star formation
followed by accretion of small satellites. They do not experience a
major merger. The simulated galaxies are old ($\approx$ 10 $Gyrs$)
hot stellar systems with masses of $\approx 1.7 \times 10^{11} 
M_{\odot}$.  Baryonic matter dominates the mass in the luminous part
of the galaxies up to $\approx 5$ effective radii. The projected properties 
of the galaxies have been investigated in detail: The \LCDM galaxy is
a slowly rotating ($(v/\sigma)_{\mathrm{max}} = 0.2$) spheroidal
stellar system (E2) with predominantly disky isophotes. The
line-of-sight velocity distributions (LOSVDs) deviate from Gaussian
shape and $h_3$ is anticorrelated with $v_{los}$. The corresponding
\LWDM galaxy is more elongated (E3 - E4) and rotates faster
($(v/\sigma)_{\mathrm{max}} = 0.6$). The  anisotropy parameter
 $(v/\sigma)^*$ is close to unity indicating 
 isotropic velocity dispersions. There is no clear indication for
 isophotal deviations from elliptical shape and the projected LOSVDs
 do not show correlated higher order deviations from Gaussian
 shape. Within the uncertainties of $M/L$ both galaxies follow the 
Fundamental Plane. We conclude that the properties of the two galaxies
 simulated in the \LCDM and \LWDM cosmology are in good agreement with
 observations of intermediate mass elliptical or S0 galaxies. Our
 conclusion differs from  Meza et al. (2003), who find, from a similar
 simulation, a much more concentrated galaxy than is generally
 observed. The differences in our findings may either be the result of
 differences in the star formation algorithms or due to the different
 merger history of the galaxies.  
\end{abstract}


\section{Introduction}
The concordance \LCDM paradigm (a cold dark matter cosmology with the
addition of a cosmological constant) appears to provide an excellent
fit to astronomical observations on scales large compared to the sizes
of individual galaxies (\citealp{2002MNRAS.337.1068P};
\citealp{2003Astro-Ph..0302209}).  However there are some indications
that the standard model may have too much power on small scales to be
consistent with observations.  For example, \cite{1999ApJ...522...82K}
show that numerical simulations predict a larger number of Galactic
satellites than observed, though \citet{2002MNRAS.333..156B} argue that this
problem can be solved by the suppression of dwarf galaxy formation in
a photoionized inter-galactic medium.  A second problem relates to the
steep cusps found in the centres of simulated dark matter halos (see
\citealp{1999MNRAS.310.1147M}; \citealp{2001ApJ...554..114E};
\citealp*{2001MNRAS.327L..27B}), which appear to be inconsistent with the
dark matter distributions inferred in dwarf galaxies (see {\it e.g.}
\citealp{astro-ph/0310001}). It is not yet clear whether this discrepancy
requires a revision of the \LCDM model. For example,
\citet{2003MNRAS.344.1237R} argues that low mass haloes in the CDM
model may have less cuspy profiles than higher mass haloes, though
this result is disputed by \citet{astro-ph/0308348}.

It is also not yet clear whether the properties of real galaxies can
be explained by the \LCDM model. Does the concordance model produce
galaxies of the right masses and sizes at the right epochs?  In fact
there are some indications from galaxy formation that the concordance
model may have too much small scale power. For example, it has proved
difficult to make realistic disk galaxies in numerical simulations of
the CDM model incorporating gas dynamics. In most simulations, the
disk systems that form are smaller, denser and have much lower angular
momenta than real disk systems (see \citealp{1994MNRAS.267..401N},
\citealp{1995MNRAS.275...56N};
\citealp{1997ApJ...478...13N}, \citealp{1999ApJ...513..555S};
\citealp{LJWPhD}). More acceptable fits to real disk systems can be
found if heuristic prescriptions modelling stellar feedback are
included in the simulations (\citealp{1998MNRAS.300..773W},
\citealp{2002Ap&SS.281..519S},\citealp{2002Astro-Ph..0207044},
\citealp{2003ApJ...591..499A}). However, even in these simulations,
the disk systems typically contain denser and more massive bulges than
the vast majority of real disk galaxies.

Most of the previous work on the formation of individual galaxies from
cosmological initial conditions has focused on the formation of disk
galaxies. The formation of individual elliptical galaxies has not been
investigated as extensively. This seems surprising as giant elliptical
galaxies are the oldest and most massive stellar systems in the
Universe and probably contribute over 50\% of the total stellar mass
if we include the stars in the bulges of S0, Sa and Sb
spirals. Although their internal kinematics can be very complex the
major component of the stellar population in ellipticals is old and
homogeneous. They are therefore good probes of galaxy assembly, star
formation and metal enrichment in the early universe (see
e.g. \citealp{2002Ap&SS.281..371T}). Furthermore, the giant
ellipticals follow simple scaling relations, the Fundamental Plane
being the most important (see e.g. \citealp{1992ApJ...399..462B}).
These simple scaling relations should arise naturally from the correct
cosmological model.

Despite their complex
kinematics, it has become evident over the last 15 years that observed
giant ellipticals show detailed photometric and kinematic
properties that correlate with their luminosity. Massive giant
ellipticals are slowly rotating, flattened by anisotropic velocity
dispersions and show predominantly boxy isophotes. Lower mass giant
ellipticals have disky isophotes and are flattened by rotation
\citep{1988A&AS...74..385B}. These low mass ellipticals most likely
contain weak disk components \citep{1990ApJ...362...52R}.  The fact
that boxy ellipticals, in contrast to disky ellipticals, show strong
radio and X-ray emission \citep{1989A&A...217...35B} and have flat
density cores \citep{1997AJ....114.1771F} might indicate that they
formed either by a different process or in a different environment.   

How and when giant ellipticals have formed is still poorly
understood. According to the ``merger hypothesis'' early type galaxies 
formed by mergers of disk galaxies. Idealised models of binary
mergers of disk galaxies (with and without gas and star 
formation) and multiple mergers have been investigated in great detail
by several authors (e.g. \citealp{1983MNRAS.205.1009N,
1992ApJ...400..460H, 1988ApJ...331..699B, 1996ApJ...471..115B,
1996ApJ...464..641M, 1996ApJ...460..101W, 2000MNRAS.312..859S}). Those
simulations  -- the recent ones with high  numerical resolution
-- are  useful in understanding detailed internal processes
{\it e.g.} gas inflow to the centre (\citealp{1996ApJ...464..641M}). They are
also capable of explaining the origin of fine structure in individual
ellipticals. For example, a large study of collisionless disk mergers
by \citet{NB2003} showed that binary disk mergers can successfully
reproduce global kinematic and photometric properties of low and
intermediate mass giant ellipticals. The formation of faint embedded
disks that are observed in these galaxies can be explained if gas was
present in progenitor galaxies (\citealp{2001gddg.conf..451N,
2001ApJ...555L..91N,2002MNRAS.333..481B}).      

Despite these successes, binary merger simulations suffer from certain
limitations. In particular they use approximate equilibrium models of
{\it present day} spiral galaxies as progenitors rather than
self-consistently calculating the properties of the progenitors ``{\it
ab initio}'' from realistic cosmological initial conditions. This is a 
serious limitation, since it is unlikely that the high redshift
progenitors of ellipticals really resembled present day spirals. Added
to the model uncertainties there are also a large number of degrees of
freedom in the initial conditions, {\it e.g.} the geometries of the
orbits, halo profiles, bulge masses, bulge rotation, gas content, gas
distribution, disk sizes {\it etc.} are all adjustable
parameters. Although there have been attempts to survey {\it e.g.}
different halo profiles (\citealp{1996ApJ...462..576D}) or disk spin
orientations (\citealp{NB2003}), these parameter surveys are evidently
incomplete.  In addition, questions regarding a self consistent
evolution of stellar populations are extremely difficult to address.
Specifically current generation ellipticals are far too red, metal
rich and old to have formed via mergers of systems similar to current
epoch spirals. Merger simulations have therefore failed, so far, to
explain the origin of global scaling relations like the
color-magnitude relation or, more generally, the fundamental
plane. They have, however, proved very useful in developing an
understanding of the detailed internal merger dynamics.
    
The best way to overcome these problems is via high resolution
simulations of individual elliptical galaxies from realistic
cosmological initial conditions. The initial conditions are
constrained by the cosmological model alone and the subsequent
evolution is governed solely by the numerical resolution and accuracy
of the physics that is implemented in the simulation. Once a
sufficient number of individual ellipticals over the whole mass
spectrum have been simulated, it should be possible (if the
cosmological model is correct) to explain the origin of the global
scaling relations and the detailed properties of individual galaxies
at different luminosities.

A first attempt in this direction has been made by
\citealp{2003ApJ...590..619M}. These authors followed the formation of a single
spheroidal galaxy in a \LCDM cosmology. The kinematic properties of
their galaxy resembled a rotationally supported giant
elliptical. However, the effective radius of the simulated galaxy was
a factor of 10 smaller than for observed ellipticals at the same
brightness. This galaxy is therefore much too compact to be consistent
with observations.

Does this result imply a problem in forming elliptical systems in the
\LCDM model?  Two lines of investigation are
suggested.  First, the cosmological model adopted may be correct but
the physical treatment may be inaccurate.  Specifically, feedback from
some early star formation into the shallow potential wells in the
small halos existent at those times, may so efficiently blow out other
gas as to reduce early star formation effectively
(\citealp{1986ApJ...303...39D}, \citealp{2000MNRAS.317..697E},
\citealp{2003MNRAS.339..289S} and Nagamine, Cen and Ostriker (in
preparation) are also exploring this possibility).
The inclusion of stellar feedback would reduce the number of low mass
galaxies, but not the number of low mass halos. It would also
significantly reduce the stellar density in the centres of systems,
but would not reduce the dark matter density by very much \footnote{A
small reduction would occur since concentration of baryonic material
forces a moderate increase of the central dark matter density by
purely gravitational processes over what would have been the case
without efficient baryonic cooling.}. 

If, however, it is not obvious that the discrepancy found by
\citealp{2003ApJ...590..619M} can be cured by a better treatment
of the physics. If the evidence for low dark matter densities in the
centres of galaxies is  taken seriously
(\citealp*{2001MNRAS.327L..27B}), then we may want to consider
a more radical solution.

An alternative is to reduce the small scale power in the dark matter
density fluctuations.  This can be achieved in various ways.  First,
the spectral index, $n$, could be sufficiently small so that after
normalization to the WMAP amplitude, and extrapolation to the small
wavelengths relevant to galaxy formation, the amplitude is low enough
to significantly reduce early star formation.  The WMAP analysis
(\citealp{2003Astro-Ph..0302209}) combined with 2dFRGS and supernova
data in fact indicated that $n$ may be as small as $0.93 \pm 0.03$,
and this may alleviate some of the purported difficulties of the
concordance \LCDM model.  We will return to this possibility in later
work.  However, there are significant limits on the value of  the spectral
index $n$, since information on the cluster length scale $\left( \sim
8h^{-1}\, Mpc
\right)$, which is intermediate between the WMAP scale and the galaxy
formation scale seems to require a relatively high normalization
(\citealp{2002AAS...201.2301B}).  This would not permit a constant $n$
solution with $n$ much less than $0.95$. Furthermore, the high optical
depth for electron scattering becomes very difficult to achieve with a
low spectral index (\citealp{2003Astro-Ph..0303236}; \citealp{2003Astro-Ph..0304234}). 

Another possibility is to achieve  low power on small scales by
some form of cut--off in the power spectrum.  An example of this is
the Warm Dark Matter model, where a finite but quite small initial
``thermal'' velocity dispersion sharply truncates the power
$\textrm{as}\, \left( k/k_{cut} \right)^{-10}$ above some wave number
scale, $k_{cut}$. \citet*{2001ApJ...556...93B} show that 
\beqn
    k_{cut} &=& 17.94 \left( \frac {\Omega _x}{0.3}\right) ^{0.15} \times \nonumber\\
    &{}&\left( \frac{h}{0.65}\right) ^{-1.3}\left( \frac{keV}{m
      _x} \right) ^{-1.15} h Mpc^{-1} ,
  \label{kcut}
\eeqn
where the warm dark matter particle has mass, $m_x$ and the density in
WDM is represented by $\Omega _x$, at $z=0$.  The additional velocity
dispersions are distributed as 
\beglet 
\beq
f(\upsilon)= \left( e^{v/\upsilon_0} +1 \right) ^{-1}\label{veldisp1} 
\eeq 
with
\beq
\upsilon _0 = 0.012 (1+z)\left( \frac{\Omega _x}{0.3} \right)
^{\frac{1}{3}} \left( \frac{h}{0.65} \right) ^{\frac{2}{3}} \left(
  \frac{keV}{m_x} \right) ^{\frac{4}{3}} km/s .\label{veldisp2} 
\eeq 
\endlet 
\citet*{2001ApJ...558..482B} used the abundance of small column
density lines in the Lyman alpha forest to limit $m_x  \approxgt 0.75 \keV$. Other
work suggests similar limits of  $m_x \approxgt 1 \keV$,
hence, in this paper, we will investigate WDM with $m_x = 1 \keV$.

One may ask at this point if the WMAP observations of a high optical
depth to the surface of last scattering, $\tau_{es}$, rule out WDM?
The answer is ambiguous.  An examination of Fig. 5 of \citet{2003Astro-Ph..0302209} indicates for $n=0.95$, $\tau_{es}$ may be $0.05$ at $1 \sigma$
level and as small as $0.01$ at the $2 \sigma$ level. Further would is
needed (but see
\citealp{2003Astro-Ph..0303622} for a counter argument) to set a limit
on $m_x$ based on WMAP. Detailed work in progress by Ricotti \&
Ostriker (2003) indicates that, if other parameters are held constant,
a WDM model with $m_x =1.25$ $KeV$ leads to only a $\sim$ 10\%
reduction in $\tau_{es}$.

In this paper we aim to investigate formation and evolution of
intermediate mass giant galaxies in the \LCDM model, and to quantify
the effect of reducing the power at small scales by studying a \LWDM
model. The paper is organised as follows: Section
\ref{Sims}, summarizes the simulation code and  describes how 
\LCDM and \LWDM initial
conditions were generated . The results of two high resolution
simulations in the \LWDM and \LCDM universe and a comparison of their
global properties with a set of low resolution simulations are
described in Section
\ref{lwcomp}. In Section \ref{obs} we compare the internal properties of the
two high resolution simulations in detail with observations of giant
elliptical galaxies. Section \ref{concs} contains a summary and our
conclusions.

\section{Initial conditions and simulations} \label{Sims}
\begin{figure*}
\begin{center}
  $\begin{array}{c}
    \hspace{-7.5cm}\mbox{(a)}\\
    \epsfig{file=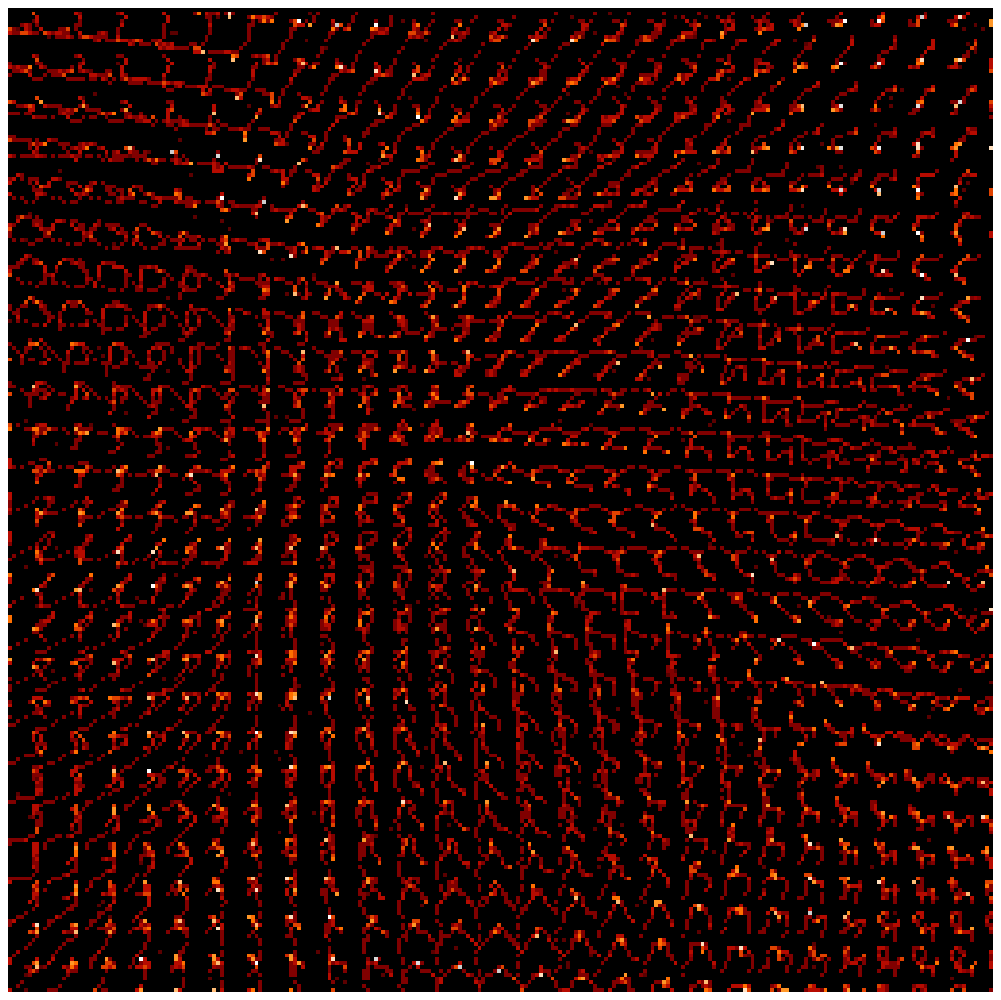,width=0.3\textwidth}\\
    \hspace{-0.5cm}\epsfig{file=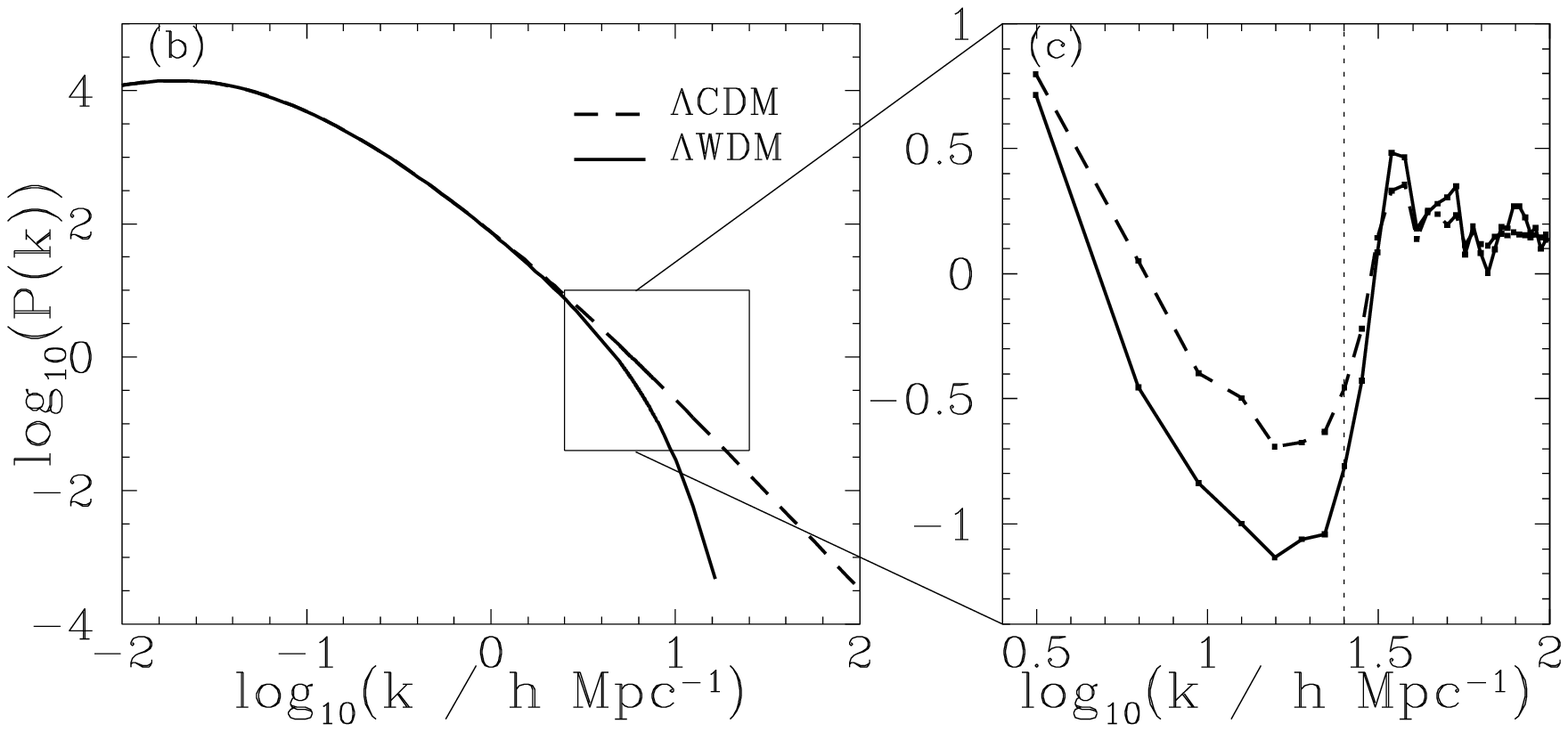, width=0.6\textwidth}\\
  \end{array}$
  \caption{{\bf (a)} The particle
    distribution of the \LWDM simulation initial conditions.
    {\bf (b)} A comparison between the input 3-D power spectrum for the
     \LCDM cosmology and a warm dark matter
    power spectrum with particle mass, m$_{x}$ = 1.0 keV. {\bf (c)} The
    calculated 3-D power spectrum from the initial conditions of the
    \LCDM and \LWDM simulations.  The dotted line is the
    Nyquist frequency of the particle grid in the case of the $50^3$
    particle simulations and is also
    the wavenumber at which noise dominates the spectrum as a result
    of the discrete nature of the initial conditions.  We note that
    the \LWDM power spectrum has a cut off which occurs at wave-numbers
    lower than this numerical limit.
    \label{Pkwdm}}
\end{center}
\end{figure*}

We performed two sets of cosmological simulations
using \LCDM and \LWDM initial conditions: 5 simulations with
$2 \times 34^3$ gas and dark matter particles and 1 simulation with $2
\times 50^3$ gas and dark matter particles.   The lower resolution simulations 
were used to investigate  the differences in the global properties of the 
stellar systems formed in each cosmology. The higher resolution simulations were
used to investigate the internal properties of the simulated galaxies.

\subsection{Power Spectra \& Cosmological Parameters}

The initial conditions of the \LCDM and \LWDM simulations
assumed scale-invariant adiabatic fluctuations.
The post-recombination power spectrum for both cosmologies is based on
the parameterisation of \citet*{1992MNRAS.258P...1E} with $\Gamma$=0.2
and the transfer function  
\beglet
\beq
T(k)=\frac{1}{\{1+[ak+(bk)^{3/2}+(ck)^2]^{\nu}\}^{1/\nu}},
\eeq 
where 
\beq
a=\left(\frac{6.4}{\Gamma}\right),
\hspace{0.2cm}b=\left(\frac{3.0}{\Gamma}\right),
\hspace{0.2cm}c=\left(\frac{1.7}{\Gamma}\right) 
\eeq
with units of h$^{-1}\Mpc\ $ and
\beq
\nu=1.13 \hspace{0.2cm} and \hspace{0.2cm} \Gamma=0.2 .
\eeq
\endlet

The power spectrum of the dark matter fluctuations in the \LCDM 
cosmology is given by 
\beq
P(k)\propto{T^2}(k)k.
\eeq 

The WDM spectrum is truncated at small scales as described by
\citet*{2001ApJ...556...93B} through an additional factor in the
transfer function of the CDM spectrum such that   
\beq T_{\rm WDM}(k)=\frac{1}{[1+(\frac{k}{k_{cut}})^2)]^5}, \eeq with
$k_{cut}$ given by Equation \ref{kcut}.  Here, $m_x$ is the mass of
the warm dark matter particle which is set to 1 \keV\  as described in 
the Introduction.   The \LWDM  power
spectrum P(k) therefore takes the following form: 

\beq P(k)\propto
T^2_{\rm WDM}(k)T^2(k)k.  
\eeq 

The amplitude of the mass fluctuations in both cosmologies is
normalised so that the $rms$ mass fluctuation in spheres of radius
8$h^{-1}$Mpc is $\sigma_8$=0.86.  This normalization was chosen to
match the present day abundances of rich galaxy clusters in this
cosmology \citep{1996MNRAS.281..703E} and is somewhat higher than the
value deduced from more recent analyses
(\citealp{2002Astro-Ph/0210567}, \citealp{2002Astro-Ph/0111362} and
\citealp{2002ApJ...569L..75V}).
 However, the value of $\sigma_8=0.86$ is consistent with the WMAP
determination and with weak lensing constraints \citep{Refrigier}, for
our adopted value of the spectral index $n=1$.

The baryonic 
fraction is set to $f_{b}= \Omega_b/\Omega_m$=0.17, consistent with
the predictions of primordial nucleosynthesis
\citep{1998ApJ...499..699B} and WMAP ($f_b=0.171\pm0.025$) 
for a Hubble parameter of h=0.65\footnote{h is defined such that
  $H_{0}$=100h kms$^{-1}$Mpc$^{-1}$.}..  Figure \ref{Pkwdm}(a) shows a
  2-d projection of a the initial particle positions for our $50^3$
  simulation in the \LWDM cosmology. Figs. \ref{Pkwdm}(b) and (c) show
  the power spectra of the \LCDM and \LWDM initial conditions
  calculated from the particle distributions at $z=24$.  The dotted
  line in Figure \ref{Pkwdm} (c) shows the limit of our spatial
  resolution in the $50^3$ simulations.  At scales smaller than this
  limit (to the right of the dotted line) there is no input power
  and the Poisson noise dominates.
\begin{table*}
\begin{center}
\begin{tabular}{|c||c|c|c|c|c|c|}
\hline  Simulation &  Number  & $\epsilon_g$ (kpc) &  
$\epsilon_d$ (kpc) & $v_c$ ($\kms$) &  $m_{dm}$ ($10^6 M_\odot$)&$m_g$
($10^6 M_\odot$)  \\ 
\hline \hline\LCDM $34^3$ & $5$    & $0.8$ & $1.4$  & 126 -- 167 & 116
-- 208 & 29 -- 52 \\ 
\hline       \LCDM $50^3$ & $1$ & $0.5$ & $0.85$ & 165 & 66 & 16 \\ 
\hline \hline\LWDM $34^3$ & $5$    & $0.8$ & $1.4$  &125 -- 169 &  116
-- 208  & 29 -- 52 \\ 
\hline       \LWDM $50^3$ & $1$ & $0.5$ & $0.85$ & 162 & 66 & 16 \\ 
\end{tabular}
\caption{Summary of some of the properties of the \LCDM SPH simulations that 
  have been re-simulated in the \LWDM cosmology. The first column gives
  the number of gas particles within the high resolution cube (equal
  to the number of dark matter particles within this volume). The
  second column lists the number of simulations at each mass
  resolution. The third and fourth columns list the Plummer softening
  parameter for the gas and dark matter particles respectively. The
  fifth column lists the range of the circular speeds of the dark
  matter haloes (determined within the virial radius at $z=0$ from the
  low resolution $AP^3M$ simulations). The sixth column lists the
  range of gas particle masses. \label{table1} }
\end{center}

\end{table*}

Velocity dispersions were added to the initial conditions of the warm
dark matter model according to equations (\ref{veldisp1} \&
\ref{veldisp2}). Additional velocity dispersions smear out the
small-scale perturbations through free-streaming.  We have performed
simulations both with and without additional velocity dispersions and
find only small differences at the very centres of the dark matter
haloes (slightly less steep cusps when the velocity dispersions were
included). The effects of initial velocity dispersions are unimportant 
for the simulations described in this paper.

We used identical values of  $\Omega_0$=0.3, $\Lambda_0$=0.7 for both
the \LWDM and the \LCDM simulations. The  mass of the warm dark matter
 corresponds to a mass cut-off of 
\beq 
M_{cut}=1\times10^{10}\Msun\left(\frac{\Omega}{0.3}\right)^{-0.45}\left(\frac{h}{0.65}\right)^{3.9}\left(\frac{m_x}{keV}\right)^{-3.45}   
\eeq
in the \LWDM simulations.
The mass resolution of the 50$^3$ \LWDM simulation is already sufficient
to correctly represent the input power spectrum above this mass cut-off (see Table
\ref{wdm:params}), thus the distribution of dark matter would not
be expected to change in a simulation with still higher resolution. 
However, this is not true of the \LCDM simulations, which have no mass cut-off.
Higher mass resolution in the \LCDM would be expected to produce  earlier
collapse of smaller structures.  

\subsection{Halo Selection}
\begin{table*}
\begin{center}
\begin{tabular}{|l||cc||cc||cc||cc||cc||}
\hline & \multicolumn{2}{|c|}{a} & \multicolumn{2}{|c|}{b}  &
\multicolumn{2}{|c|}{c} & \multicolumn{2}{|c|}{d} &
\multicolumn{2}{|c|}{e}\\ 
\hline & \LCDM & \LWDM & \LCDM & \LWDM & \LCDM & \LWDM & \LCDM & \LWDM
& \LCDM & \LWDM \\ 
\hline \hline {R$_{vir}$/kpc} & 350 & 360 &330 &330& 280 & 280 & 280 & 280&300 & 310\\
\hline M$_{DM}$(R$_{vir}$)/10$^{11}$M$_{\odot}$ & 21 & 22&17 & 17&10 &10&10 &11&13 &14\\
\hline log$_{10}\left(j_{h} \left( R_{vir}\right)\right)$ &
3.1& 3.1 & 3.2& 3.1 &3.3& 3.4 &3.3& 3.3 &3.2& 3.3\\ 
\hline M$_{gal}$/10$^{10}$M$_{\odot}$ & 8.0& 8.9&7.3 & 7.9 & 7.4& 6.1
&5.5&6.9 &6.3& 7.2\\ 
\hline log$_{10}\left(j_{*}/\kpc \kms\right)$ & 1.7 & 2.3 &2.1& 1.2
&1.4& 1.5 &2.6& 2.7 &2.0& 2.2 \\ 
\hline
\end{tabular}
\caption{The virial radii, virial masses, galaxy masses and
   specific angular momenta of the galaxies (measured in $\kms
  \kpc$) and dark matter haloes for each of our five, $34^3$ simulations in the \LCDM
  and \LWDM simulations at z=0.  The difference between the cold and warm
  dark matter simulations are not statistically significant at the
  gross, integrated level shown in this table.}\label{wdm:params}
\end{center}
\end{table*}

The initial conditions for our simulations were generated in a two
stage process.  A low resolution dark matter only simulation of a
large computational volume was run until the present day. Haloes were
selected from the final output of the simulation to be run at higher
resolution including Smooth Particle Hydrodynamics (SPH) to represent
a dissipative gas component.  The re-simulation of these haloes at
higher resolution requires the addition of short wavelength
fluctuations missing in the low resolution simulations. In addition
particles at large distances are averaged hierarchically to represent
the tidal field of the entire computational volume by a manageable
number of particles. A full description of the methods used to
generate our initial conditions can be found in
\cite*{1998MNRAS.300..773W} (see also \citealp*{1997ApJ...490..493N}).
Here we summarise the specific parameters used for the simulations in
this paper.

We ran a low resolution dark matter simulation using the Adaptive
Particle-Particle, Particle-Mesh (AP$^3$M) N-body code of
\cite{1991ApJ...368L..23C}.  For both cosmologies, a cube of
$L_{box}$ = 50 \Mpc  containing 128$^3$ particles was evolved from a
redshift of z = 24 to the present day. The force law used in the code
was a Plummer law with a gravitational softening  fixed in
comoving coordinates at 7 \kpc. Identical random phases were used in
both cosmologies so that the same haloes could be simulated at higher
resolution enabling direct comparisons to be made.
We used a spherical over-density group-finding algorithm
(\citealp*{1994MNRAS.271..676L}) to identify virialised haloes at
$z=0$ with masses in the range of $2\times 10^{12} M_{\odot} < M_{halo} <
7\times 10^{12}$ in a low density environment such that the nearest
halo with a mass greater than  $7.1 \times$ 10$^{11}M_{\odot}$ is
over 1 \Mpc\ away.  If applied to a 50$^3$ $Mpc^3$\ volume of the
real universe our selection procedure would identify
intermediate-mass giant field ellipticals or S0 galaxies systems
similar to the  Sombrero galaxy (M104) rather than late type
spirals which are of lower mass,  or very bright giant ellipticals which
are found in high density regions.     

Each target halo was re-simulated at a higher resolution with the
inclusion of a gas component. We increased the particle number within
a cubic volume at redshift $z=24$ containing all particles that end up
within the virialised region of the halo at $z=0$. Additional short
wavelength perturbations were included to account for the missing
small-scale power below the Nyquist frequency of the low resolution
simulation.  The size of the high resolution cube was of the order of
$L_{box} = 5 kpc$. The tidal forces from particles outside the high
resolution cube were approximated by increasingly massive particles in
5 nested layers of lower and lower resolution. Gas particles with
masses chosen to the match the baryonic fraction were added  in
the high-resolution regions,  in the same number and at the same
positions as the dark matter particles. Outside the high-resolution
regions only dark matter particles were included. 

In total we ran 6 simulations simulations from $z=24$ to $z=0$ for
each of the \LWDM and \LCDM cosmologies: Five simulations with
$2\times 34^3$ particles (labelled (a) to (e)) and one simulation with
$2 \times 50^3$ particles within the high resolution cube . The
simulations will be referred to as the $34^3$ and $50^3$ simulations
hereafter. At the mass and spatial resolution of the $34^3$ runs, it
is not possible to resolve the internal structure of a forming galaxy
and they are used instead to investigate variations in their global
properties. The $50^3$ simulations are used to investigate the
internal properties of the galaxies in detail.
\begin{figure}
\begin{center}
        \epsfig{file=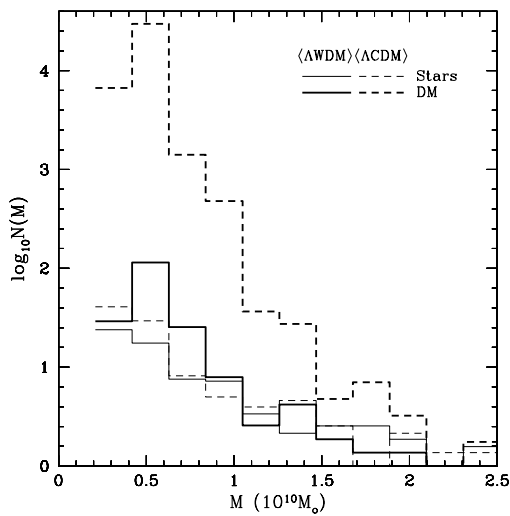,
        width=0.4\textwidth} 
        \caption{The number of groups (averaged over the 5 simulations) of a given 
          mass in the 34$^3$ \LWDM (solid) and \LCDM (dashed) simulations for both
        dark matter (thick) and stars (thin) at $z=0$.
          \label{histogram}}
\end{center}
\end{figure}
   
The simulations were performed with the GRAPESPH code
as outlined in \cite{1998MNRAS.300..773W} using 5 GRAPE-3A boards
(\citealp{1990Natur.345...33S}) connected to a Sun Ultra-2
workstation. The evolution of the gas component was followed with SPH
including radiative cooling and a simple star formation algorithm:
Each gas particle that remains in a collapsing region with a density  
$\rho_{crit}{>}$7x10$^{-23}$ \kg \m$^{-3}$ (see
\citealp*{1993MNRAS.265..271N}; \citealp{1998MNRAS.300..773W}) for
longer than a local dynamical time was converted into a star particle.   
Stellar feedback has been included in some simulations,
using somewhat `ad hoc'  rules (\citealp{2002Ap&SS.281..519S}; 
\citealp{2003ApJ...590..619M}; \citealp{2003MNRAS.339..289S}). However, it
is clear that these representations are far from realistic. We have
therefore chosen to make our simulations as simple as possible and
hence stellar feedback is ignored. This makes it easier to compare
simulations done at different mass resolutions, and to understand
if the properties of the final stellar systems depend on whether
the dark matter is warm or cold. Theoretical arguments (see {\it e.g.}
\citealp{1986ApJ...303...39D}; \citealp{2000MNRAS.317..697E}) suggest that stellar
feedback is less important in high velocity dispersion systems than in
low velocity dispersion systems, though since galaxies form
hierarchically,  feedback processes may have a significant effect on the
evolution of high velocity dispersion systems at early times.

\section{\LCDM vs. \LWDM} \label{lwcomp}
\noindent

\begin{figure}
\begin{center}
  \epsfig{file=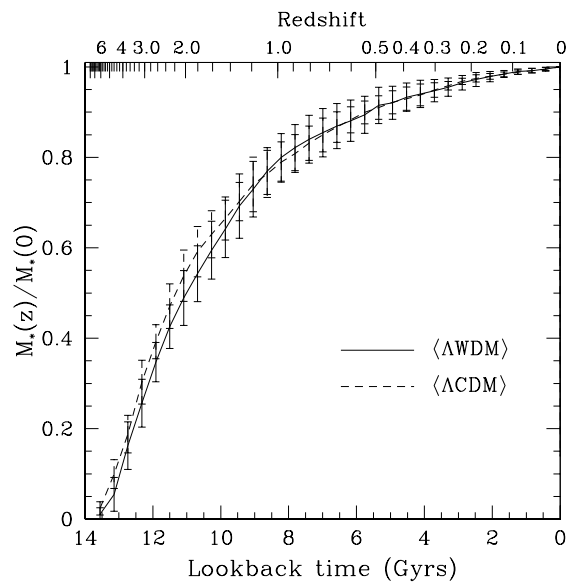, width=0.45\textwidth}
  \caption{Normalised stellar mass of the central object
    versus lookback time, averaged over the 5 simulations in the \LCDM
  (dashed) and \LWDM simulations (solid).  The error bars indicate the
  one sigma error.     On average the \LCDM galaxy assembles its stars
  earlier than the \LWDM galaxy.  \label{mfvz}}
\end{center}
\end{figure}

\subsection{$34^3$ Simulations}  
\begin{figure*}
\begin{center}
\epsfig{file=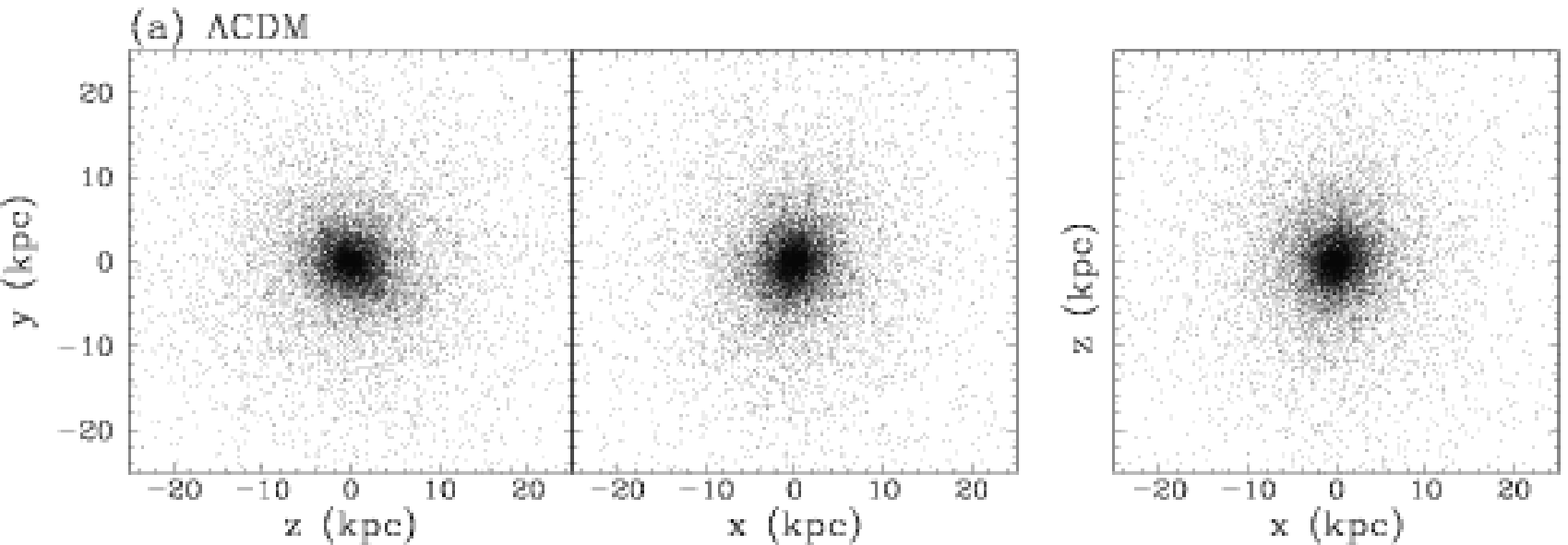, width=0.9\textwidth}
\epsfig{file=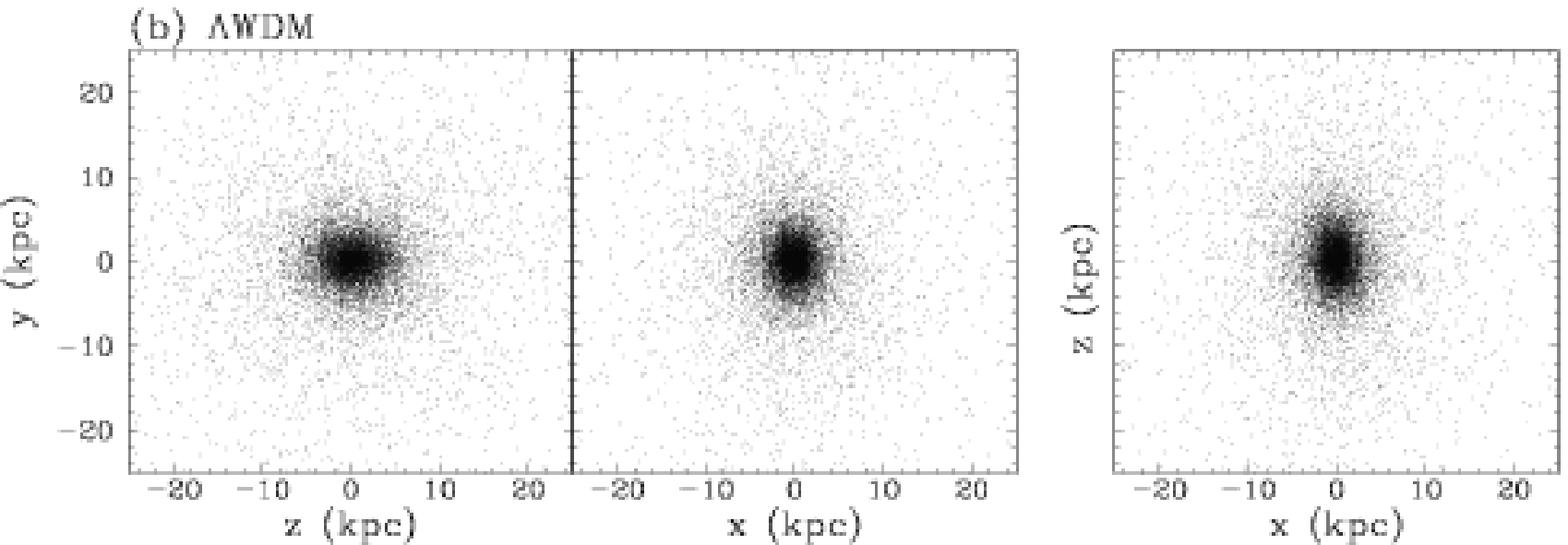, width=0.9\textwidth}
\caption{ Three orthogonal projections of the star particle positions
of the {\bf{(a)}} \LCDM and {\bf{(b)}} \LWDM galaxy in the $50^3$
simulations, at z=0. \label{poses} }  
\end{center}
\end{figure*}

For the \LWDM cosmology investigated in this paper the formation of
halos with masses $M_{halo}<M_{cut}\simeq1\times10^{10} M_{\odot}$ should
be suppressed because the initial power spectrum is truncated below
this mass-scale. To quantify this effect, we compared the mass spectrum
of satellite halos in the two cosmologies at the present day. A
friends-of-friends group finding algorithm with a linking length of
$0.8$ $kpc$ was used to identify dark matter haloes and their stellar
systems for the $34^3$ simulations at $z=0$ . The massive central
galaxy was excluded.

Figure \ref{histogram} shows the mass distributions of the dark matter
and stellar systems averaged over the 5 simulations in each cosmology
As expected, the \LCDM cosmology produces many more dark matter haloes
with masses $\leq 1\times10^{10} {\rm M_{\odot}}$. There is, however,
only a marginal difference in the mass distribution of the stellar
systems.
The central galaxies in the $34^3$ simulations have halo masses in the
 range of $1.0 \times 10^{12} M_{\odot}$ to $2.2 \times 10^{12}
 M_{\odot}$, well above the cut-off mass of the \LWDM cosmology.
 Table \ref{wdm:params} summarizes some of the global properties of
 the \LCDM galaxies are their \LWDM counterpart. The virial radii,
 masses and the specific angular momenta of the galaxies and their
 halos do not reveal any large differences between the cold and warm
 dark matter simulations.

A comparison of the assembly history of the stellar mass for the
central \LCDM and \LWDM galaxies averaged over the five $34^3$
simulations is shown in Figure \ref{mfvz}. The mean values and
standard deviations are shown. The redshifts at which ${1/10}$ and
${1/2}$ and ${9/10}$ of the present day mass in stars has been
assembled are listed in Table 3. Again, there is no obvious difference
between the \LCDM and the \LWDM cosmologies. At $z=0$ the stellar systems
in both cosmologies have mean ages of about $10$ $Gyrs$. The masses and ages
of the galaxies are thus comparable to those of early type galaxies. In
conclusion, despite differences in the formation of low mass dark
matter haloes,  the massive central galaxies formed in the two
cosmologies resemble early type galaxies and have indistinguishable
formation histories and global properties. The resolution of the
$34^3$ simulations, however, is too low to investigate the internal
properties of the main stellar systems in  detail. We therefore now
turn to the higher resolution $50^3$ simulations.

\subsection{$50^3$ simulations}\label{internals}

\begin{figure}
\begin{center}
\epsfig{file=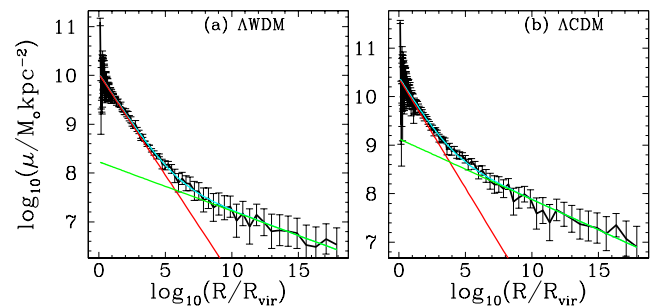, width=0.5\textwidth}
\caption{The radial density profile for the galaxies in our $50^3$ (a)
\LWDM simulation and (b) \LCDM simulation.  Two exponential fits are
also plotted in the figure to represent an inner and outer
  component. The combined fit is also shown. The best fitting 
  exponentials have been calculated according to
  Eqn. \ref{eqn:fits}.\label{surf}} 
\end{center}
\end{figure}

We have re-simulated  halo `a' (see Table
\ref{wdm:params}) using $2 \times 50^3$ particles within the high
resolution region. Figure
\ref{poses} shows the three orthogonal projections along the principal
axes of the moment of inertia tensor of the central stellar systems in
the \LCDM and \LWDM cosmologies at $z=0$. The \LCDM simulation
produces a spheroidal stellar distribution.  The \LWDM galaxy on the
other hand has a slightly more flattened morphology. We assume for a
moment that the galaxies are two component systems and perform a
decomposition of the surface mass density profile into an inner and
outer component of the form:
\beq
\log_{10}\left[\frac{\mu
    (r)}{\mu_0}\right]=\log_{10}\left[a \exp \left(\frac{-r}{b}\right)+c \exp
  \left(\frac{-r}{d}\right) \right], \label{eqn:fits} 
\eeq
where the constants $a, b, c$ and $d$ were determined using a non-linear
fitting algorithm.  The surface density profiles and the fits for
the two galaxies are shown in Figure \ref{surf}.  
The stellar system that forms in the \LWDM simulations has a more
extended outer component.  More interestingly, however, we see
that the inner component of the galaxy in the \LWDM simulation is more
extended than in the \LCDM galaxy.
To test for evidence of whether the inner and outer components have
kinematic counterparts,  we performed a kinematic decomposition of the
stars. We used a weak criterion to separate a kinematically cold
component from a hot component. A star was tagged as a member 
of a cold component if  
\beglet \beqn
v_{\theta,i} &>& 0.7v_{c}  \label{decomp1}\\
v_{\theta,i} &>& 2 \langle{v^2_{rand}}\rangle_i^{1/2} \label{decomp2}\\
\langle{v^2_{rand}\rangle} &=&
\frac{1}{3}(v_{r,i}^2+(v_{\theta, bin}-\overline{v_{\theta,i}})^2
+ v_{z,i}^2) \label{decomp3}
\eeqn \endlet 
where $v_c$ is the average circular velocity of the radial bin
containing the star, $v_{r,i}$ is the radial velocity of the star,
$v_{\theta,i}$ is the rotational velocity of the star, $v_{z,i}$ is
the velocity along the minor axis defined by the moment of inertia
tensor and $\overline{v_{\theta,bin}}$ is the
average rotational velocity of the radial bin containing the
star. Both galaxies are essentially hot stellar systems with no
significant indication of a dynamically cold sub-component. The \LWDM
galaxy contains a small number of particles 
which obey the criteria given in equations \ref{decomp1} -
\ref{decomp3} which are, however, distributed in a thick disk-like
structure.    

\begin{figure}
\begin{center}
\epsfig{file=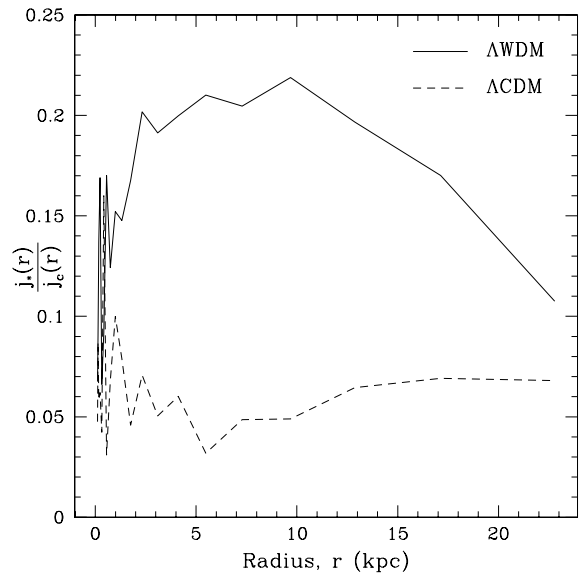, width=0.45\textwidth}
\caption{The radial profile of the stellar specific angular momentum 
  as a fraction of the specific angular momentum calculated from the
  circular velocity. The main stellar system in the \LCDM (dashed)
  cosmology has only weak rotational support.  The \LWDM (solid) galaxy has
  more angular momentum in the outer regions.\label{jmvr}} 
\end{center}
\end{figure}

To quantify whether the flattening of the \LWDM galaxy is caused by
rotation, we have calculated the dimensionless spin parameter,$\lambda$,  for the
dark matter halos and the stellar systems where $\lambda$ is defined as
 \beq
\lambda=\frac{J|E|^{1/2}}{GM^{5/2}}, \label{dimspin} \eeq and J is
the angular momentum, E is the total energy and M is the total mass of
the system (see {\it e.g.} \cite{1980MNRAS.193..189F}).  To compute
(\ref{dimspin}) only bound halo particles inside the virial radius and
bound stars within the outer radius of the luminous part of the galaxy
are included (defined as the radius at which the stellar density of
the galaxy averaged in spherical shells becomes equal to the
background level). The stellar system is 
  treated as if it were virialised and so  $|E|$ in Eqn.
\ref{dimspin} is replaced by the kinetic energy of the stars $|T_*|$).
In the \LCDM cosmology the halo has $\lambda_H$=0.041 and the stars
  have $\lambda_*$=0.053, in the \LWDM cosmology the halo has a
  slightly lower value of $\lambda_H$=0.032 whereas the stars have
  higher value of $\lambda_*$=0.228.  Fig.  \ref{jmvr} shows the
  specific angular momentum profiles for the \LCDM and the \LWDM
  galaxy. Neither cosmology produces a purely rotationally supported
  stellar system ($j_*(r)/j_c(r) \approx 1$, where $j_c(r)=rv_c(r)$ is
  the specific angular momentum computed from the circular
  speed). However, the stellar system of the \LWDM galaxy has a higher
  angular momentum than the \LCDM galaxy except in the central regions
  $r \approxlt 1  kpc$.

\begin{figure}
\begin{center}
\epsfig{file=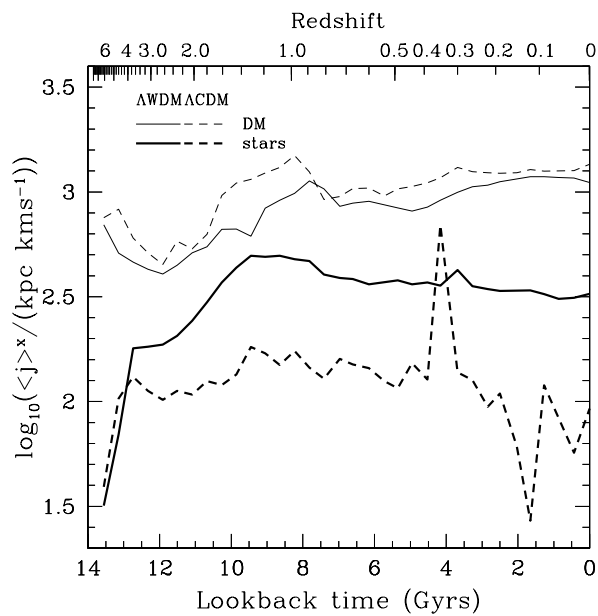, width=0.475\textwidth}
\caption{The evolution of the specific angular momentum 
  of stars (thick) and dark matter (thin) for the \LCDM (dashed) and
  \LWDM (solid) galaxy. 
  \label{jmvzwbig}}
\end{center}
\end{figure}

Figure \ref{jmvzwbig} shows the time evolution of the average specific angular
momentum of the dark matter and stars in both cosmologies. At any
given redshift the dark matter halo of the \LCDM galaxy has more
angular momentum than the \LWDM halo,  but they both show a very similar
evolution with time.  
In contrast, the angular momentum of the stars of the \LCDM galaxy
does not change significantly after a redshift of $z=4$ whereas the
stars of the \LWDM  galaxy gain further angular momentum until $z=1$
with little evolution thereafter.

\begin{figure}
\begin{center}
\epsfig{file=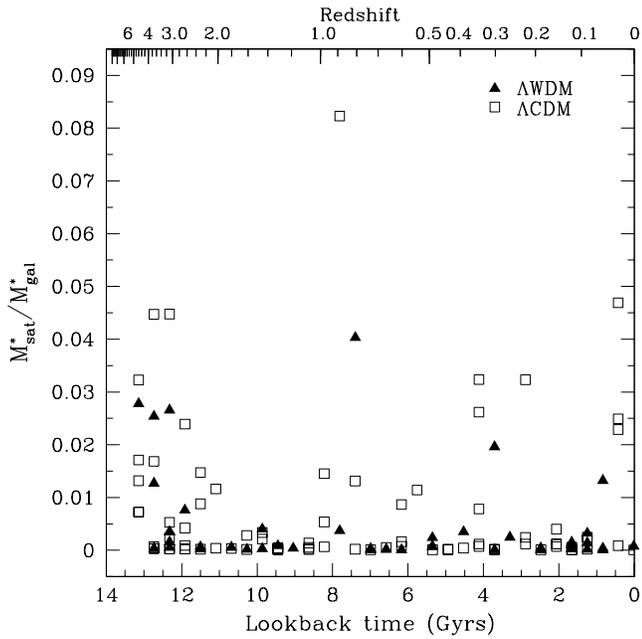, width=0.5\textwidth}
\caption{Ratio of the stellar mass of satellites that merge with the central
galaxy to the stellar mass of the galaxy as a function of lookback
  time for both the \LCDM (open squares) and \LWDM (filled triangles) cosmology. 
	Every identified satellite with a mass larger than 
         $3.2 \times 10^{18} M_{\odot}$ is plotted.  
 \label{mergers}}
\end{center}
\end{figure}

\begin{figure}
\begin{center}
\epsfig{file=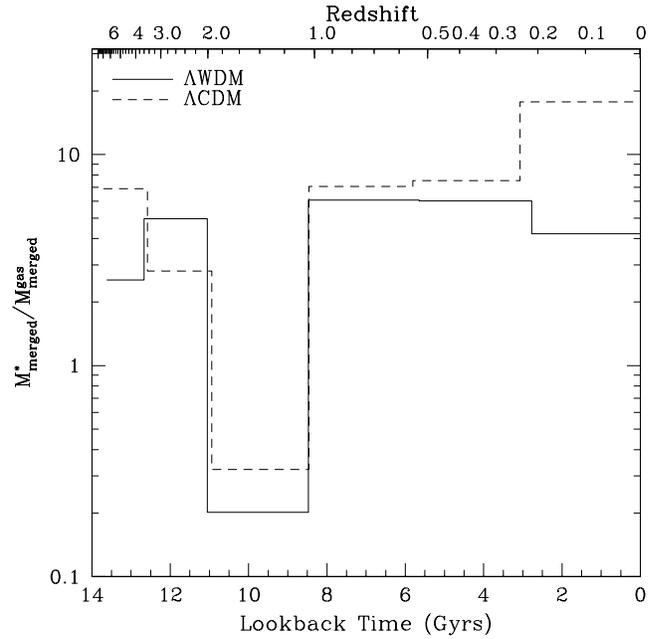, width=0.5\textwidth}
\caption{The star to gas mass-ratios of merged satellites as a
function of lookback time for the \LWDM (solid) and \LCDM (dashed)
simulations.  Satellites merging at $z<1$ are in general more gas rich
than \LCDM satellites. \label{gasfrac}} 
\end{center}
\end{figure}

\begin{figure}
\begin{center}
\epsfig{file=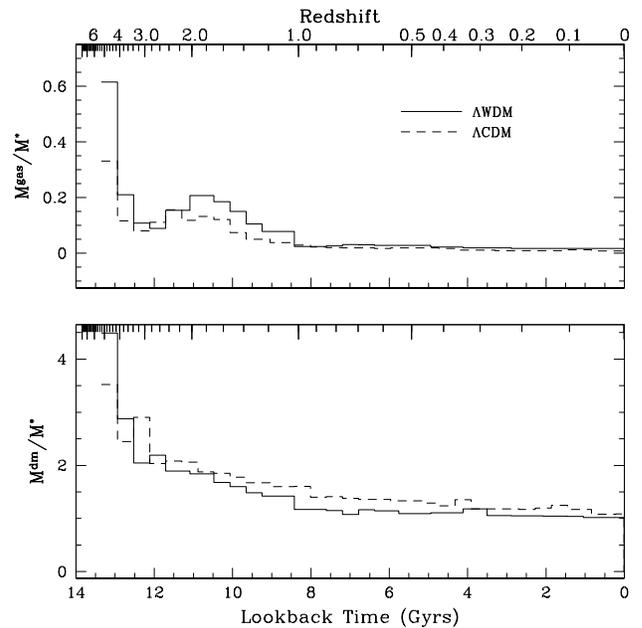, width=0.5\textwidth}
\caption{The time evolution of the mass-ratio of gas to stars (top)
and dark matter to stars (bottom) of the \LWDM (solid) and \LCDM (dashed)
galaxy within 20 $kpc$. Throughout the evolution the \LWDM galaxy is
more gas rich than the \LCDM galaxy. \label{fractions}} 
\end{center}
\end{figure}

We have investigated the merger histories of the galaxies in more
detail. A friends-of-friends algorithm with a linking length of 2.0
$\kpc$ ($\approxgt 2\epsilon$) was used to identify satellite galaxies
consisting of more than 20 stellar particles ($3.2 \times 10^8
M_{\odot}$) at each redshift. A satellite was then defined to have
merged with the central galaxy at a redshift at which its particles
were first identified as members of the central galaxy. Figure
\ref{mergers} shows the stellar mass ratio of every identified merger
as a function of lookback time. The total number of mergers with
M$_{sat}^*$ $>$ 10$^9$ M$_{\odot}$ is 58 in the \LWDM simulation
compared with 96 in the
\LCDM simulation. The results show that at almost every redshift
the \LCDM galaxy experiences more individual minor mergers with higher
mass ratios in stars than the corresponding \LWDM galaxy. The mass
ratio for an individual merger never exceeds 10:1. Low accretion
rates like this are expected as the initial conditions were 
selected from low density environments. In addition to the frequency
of mergers, the average internal composition of the merging satellites changes 
with time and is different in the two cosmologies (Figure \ref{gasfrac}). At 
redshifts $z < 2$ the \LCDM satellites always have a higher star-to-gas ratio. 
Interestingly, the only redshift range where the satellites have more mass 
in gas than in stars is $2 < z < 1$. This corresponds to the period where 
the galaxies reach their second peak in the gas-to-star mass ratios as shown 
in Figure \ref{fractions}. In this figure 
we plot the time evolution of the composition of
the central galaxy. The mass-ratios of gas to
stars (top) and dark matter to stars (bottom) in the \LWDM and  \LCDM
galaxies are plotted. The dark matter fractions vary only slightly
between the two cosmologies. However, between a redshift of $z=4$ and
$z=1$ the \LWDM galaxy is significantly more gas rich than the \LCDM galaxy.    

\begin{figure}
\begin{center}
\epsfig{file=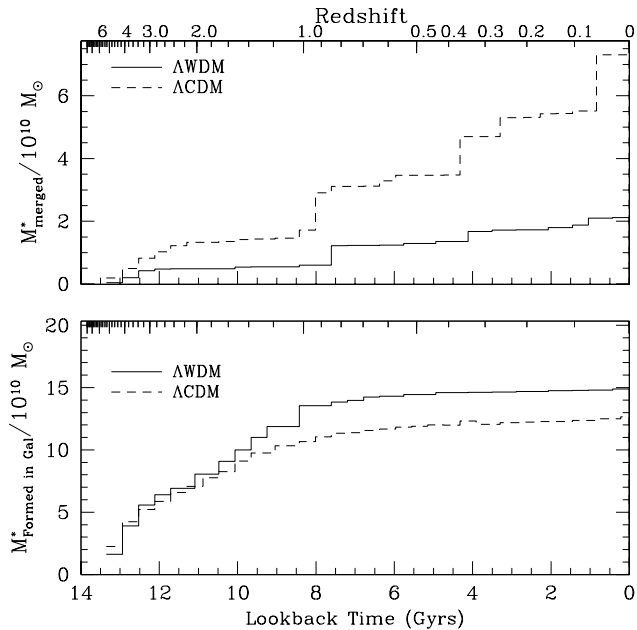, width=0.5\textwidth}
\caption{{\bf Top:} The cumulative stellar mass accreted in mergers as a
function of lookback time for the \LWDM galaxy (solid) and the \LCDM
galaxy (dashed). {\bf Bottom:} The evolution of the stellar mass formed
within the galaxies. The masses include material inside a radius of 20
$kpc$. After $z=2$ the galaxy in the \LWDM simulation accretes fewer
stars from mergers and interactions and forms the majority of its new
stars within the galaxy itself.  The \LCDM galaxy, however, accretes
nearly half its mass in stellar mergers. \label{resid}} 
\end{center}
\end{figure}

To investigate where the stars that end up in the final galaxies have
formed,  we have calculated the stellar mass accumulated by
the galaxies through mergers and compared it to the cumulative mass of
stars that formed inside the galaxies (Figure \ref{resid}). Below a
redshift of two, when both galaxies have already assembled $\approx
50\% $ of their final stellar mass (see Figure \ref{newmass}),
about 30\% of all present day stars in the \LCDM galaxy have 
been accreted by mergers and $\approx 20\%$ of the stars have formed  
inside the galaxy. Over the same period,  the \LCDM galaxy has accreted only
$\approx 10\%$ of its present day stars and $\approx 40\%$ have formed
within the galaxy.    
  
\begin{figure}
\begin{center}
\epsfig{file=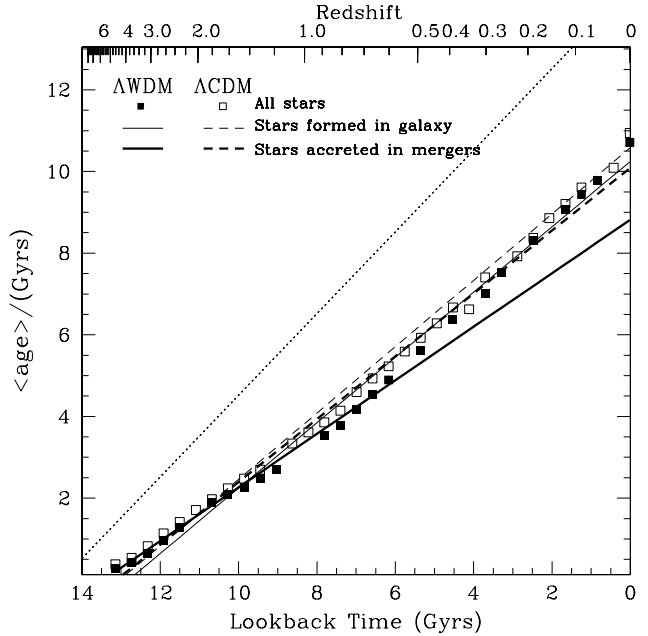, width=0.5\textwidth}
\caption{The average age of all the stars within the galaxy in the
\LWDM (filled squares) and \LCDM (open squares)  
 cosmologies. Linear least squared fits are plotted for the average
 age of stars formed in the  galaxy (thin) and the
 average age of stars accreted in mergers (thick) for \LCDM (dashed)
 and \LWDM (solid).  The
dotted line represents the age of the universe.  The stellar
populations of the mergers in the \LWDM simulation are on average
younger than the corresponding \LCDM populations.  At $z<1$ there is a
difference of approximately 1 $Gyr$ in age. \label{mergages}} 
\end{center}
\end{figure}

The influence of the different accretion histories on the mean ages of
the stellar population of the galaxies is shown in Figure
\ref{mergages}. In each redshift bin we show the average age of the
total stellar population for the \LCDM (open squares) and the \LWDM
(filled squares) galaxies. In addition, the total stellar
population was divided into stars that have merged with the
galaxy (thick dashed and solid lines) and stars that have formed
in the galaxy (thin dashed and solid lines). At all redshifts the \LCDM
galaxy accretes stars that are of a similar age to the 
galaxy itself. The accreted stars have formed in small halos at the same
time as the stars within the main galaxy.
In contrast, the stars accreted in the \LWDM universe below $z=0.5$
are on average 1 $Gyr$ younger than the stars within the galaxy. 
In total, however, the number of accreted stars is so small that there
is almost no effect on the mean age  of the \LWDM
galaxy.      

\begin{figure}
\begin{center}
\epsfig{file=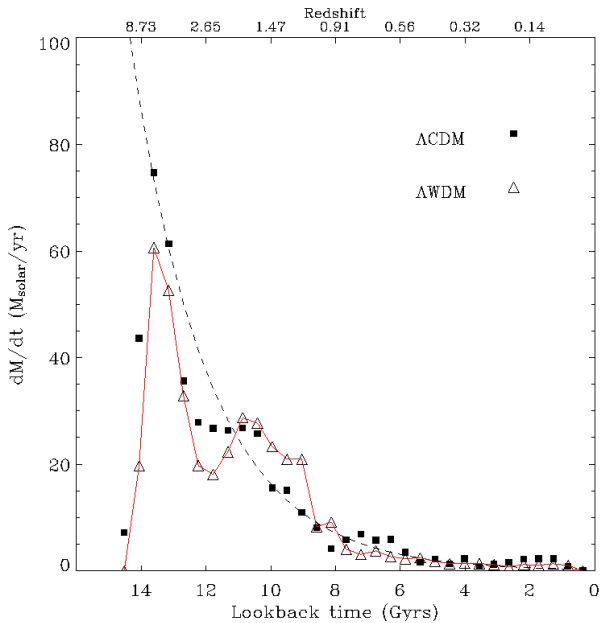, width=0.475\textwidth}
\caption{Star formation rate of the \LCDM and the \LWDM galaxy versus
  lookback time. The evolution of the \LCDM galaxy is close to
  exponential with a star formation rate $ \propto \exp{(-(t-t_0)/\tau)}$
  with $t_0=13.5\ Gyrs$ and $\tau = 2.4\ Gyrs$. The \LWDM galaxy shows a
  strong peak approx. $13.5\ Gyrs$ ago and a broad second peak a little
  more than  $10\ Gyrs$ ago. \label{sfr}}
\end{center}
\end{figure}

Fig. \ref{sfr} shows the star formation rate of the main galaxy in the
\LWDM and \LCDM simulations.  The star formation rate of the \LCDM
galaxy can well be approximated by an exponential, sfr $ \propto
\exp{(-(t-t_0)/\tau)}$, with $t_0=13.5\  Gyrs$ and $\tau = 2.4\
Gyrs$. The \LWDM galaxy does not follow an exponential. There is
evidence for a second broad peak in the star formation rate around
$10\ Gyrs$ ago which  is related to  the enhanced gas accretion
of the \LWDM galaxy at early times.    

The stellar mass assembly history of the 50$^3$ galaxies is shown in
Fig. \ref{newmass}. As for the $34^3$ simulations,  the early star
formation rate of the central galaxy in the \LWDM cosmology is
significantly lower than in the \LCDM  cosmology.  Due to this
difference in the star formation histories,  the \LCDM galaxy is
slightly older (half of its stars are older than $11.9\ Gyrs$) than
the \LWDM galaxy (half of the stars are older than $11.2\ Gyrs$). 

\subsection{Convergence Tests: Anticipated Effect of Extrapolation to
  Higher Resolution}  

In this section we investigate the effect of numerical resolution on
the global properties of the galaxies formed in the \LWDM and \LCDM 
cosmologies. We estimate the degree to which convergence has been
reached by comparing our $34^3$ and $50^3$ runs, which have identical large scale 
perturbations. 

\begin{figure}
\begin{center}
\epsfig{file=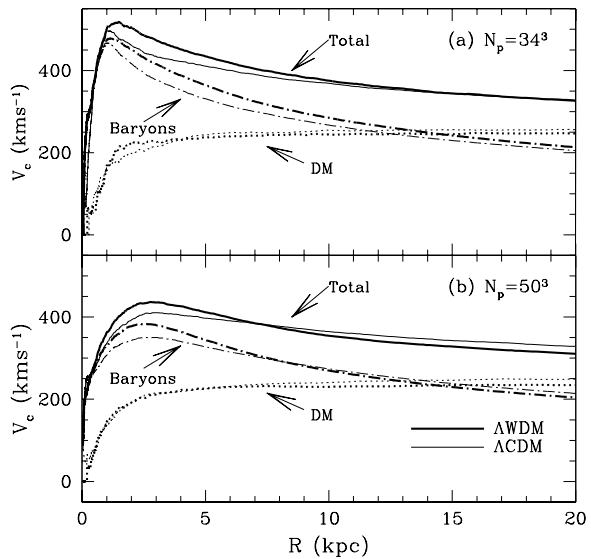, width=0.45\textwidth}
\caption{The circular velocity profiles of the $34^3$ (upper
  panel) and $50^3$ (lower panel) \LWDM and \LCDM  galaxies. 
  The total velocity curves are shown by the solid line. The
  baryonic (dot-dashed) and dark matter (dotted) contributions are plotted
  separately. The dynamics of the inner regions ($r < 10$ $kpc$) is 
dominated by baryonic material,
  which is more  concentrated at lower resolution, where as the outer
  regions of the galaxies are dominated by dark matter. \label{vcirc}} 
\end{center}
\end{figure}

Table \ref{rescomp} lists some global properties of the $34^3$ 
and $50^3$ galaxies. The only value that increases significantly with 
resolution for both cosmologies is the radius at which the circular 
velocities reach their peak values. The corresponding circular velocity 
profiles for the baryonic and dark matter are shown in Fig. \ref{vcirc}.
The baryonic matter is more concentrated at low resolution. There 
are a number of possible explanations for this behaviour. If the gas
in the inner parts of the galaxy is poorly resolved,  star formation 
leads to a rapid depletion of gas resulting in artificial 
pressure gradients which drive more gas to the center. There is also 
the possibility that a mismatch of gravitational softening 
(Plummer) and SPH softening (cubic spline) supports unstable 
regions to collapse artificially (see \citealp{1997MNRAS.288.1060B}). 

\begin{table}
\begin{center}
\begin{tabular}{|r||c|c|c|c}
\hline & \multicolumn{2}{|c|}{\LCDM} & \multicolumn{2}{|c|}{\LWDM} \\
 & 34$^3$ & 50$^3$ & 34$^3$ & 50$^3$ \\

\hline \hline R$_{max}^*$/(kpc) & 1.0 & 2.9 &1.5 & 2.7\\
\hline T$_{rot}$/T$_{rand}$ & 0.006 & 0.004 & 0.055 & 0.064\\
\hline $\lambda_*$ & 0.046 & 0.053& 0.176  &0.228  \\ 

\hline M(R$^*_{max}$)/(10$^{10}$M$_{\odot}$) & 4.49 & 7.63  &6.84 & 8.43 \\

\hline $\rho^*(R_{max})$/(10$^{10}$M$_{\odot}$kpc$^{-3}$) & 1.09 &0.08 &0.53 & 0.11\\

\hline $(1+z_{1/2}^*)$ & 3.40 & 3.62  &3.07 &3.08\\
\hline $log_{10}(Age^*(R_{max})/yrs)$ & 9.971 & 10.01 & 9.987 & 10.01\\
\hline & & && 
\end{tabular}
\caption {The physical parameters of the stellar system formed in our
\LCDM and \LWDM simulations at the $34^3$ and $50^3$
resolution. $R_{max}^*$ is the characteristic radius as calculated
from the peak in the circular velocity profile, $\lambda_*$ is the
dimensionless spin parameter for the stars, $T_{rot}/T_{rand}$ is the
ratio of the rotational kinetic energy to the energy in random
motions, $M_*$ is the total mass within R$_{max}^*$, $\rho^*(R_{max})$
is the density of the stars within R$_{max}^*$, $z_{1/2}$ is the
redshift at which half of the stars have formed and $Age(R^*_{max})$
is the average age of the stars within R$_{max}^*$. \label{rescomp} }  
\end{center}
\end{table}

\begin{figure}
\begin{center}
\epsfig{file=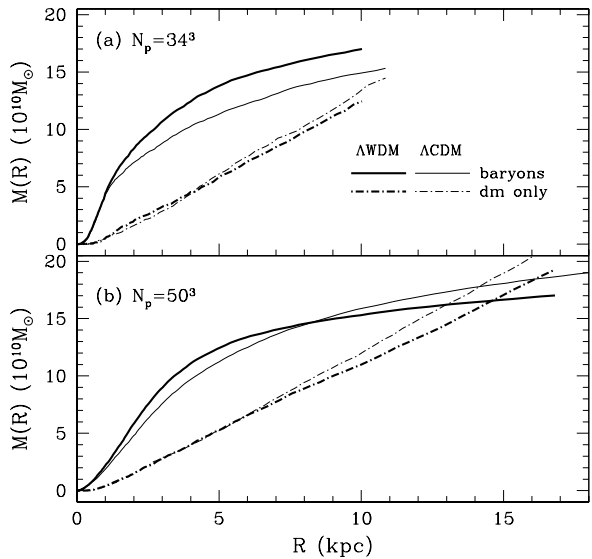, width=0.45\textwidth}
\caption{ The cumulative mass profiles for the $34^3$ (upper panel) and
  $50^3$ (lower panel) \LWDM (thick) and \LCDM
  (thin) simulations plotted for the dark
  (dot dashed) and baryonic (solid) matter separately. Within at least
  5 effective radii, the galaxies are dominated by dark matter.\label{masscomp}} 
\end{center}
\end{figure}

The cumulative mass profile is shown in Fig. \ref{masscomp}.  The
galaxies produced in the 50$^3$ simulations extend much further than
the lower resolution galaxies due the density criterion that we use to
define the size of the galaxy.


\begin{figure}
\begin{center}
\epsfig{file=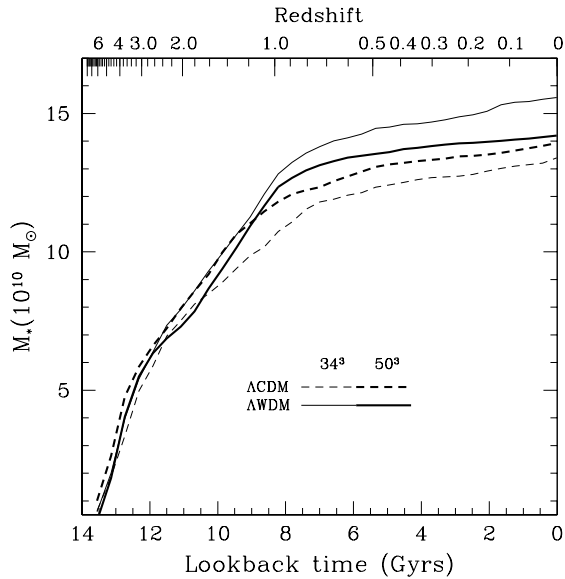, width=0.45\textwidth}
\caption{The evolution of stellar mass within 7kpc 
($\approx 2 r_{\mathrm{eff}}$) for the 
$34^3$ (thin) and $50^3$ (thick) simulations in
  the \LWDM (dashed) and \LCDM (solid) cosmologies.  The \LWDM simulation
  appears to have converged by the resolution achieved in the $50^3$
  simulation, but this is not true of the \LCDM
  simulation. \label{newmass}} 
\end{center}
\end{figure}

Fig. \ref{newmass} compares the stellar mass within 7 $kpc$,
corresponding to $\approx 2 r_{\mathrm{eff}}$, as a function of
lookback time for the $34^3$ and $50^3$ simulations in both
cosmologies. We see that the assembly history is resolution dependent
for redshifts $z < 2$. The trend goes in opposite directions for the
two cosmologies. The \LWDM galaxy  accretes more mass whereas the
\LCDM galaxy accretes less. The final masses seem to converge at
higher resolution. However, we note that our simulations do not
include a photo-ionising background which would improve the numerical
convergence in the case of the \LCDM cosmology.    

The collapse of objects smaller than the resolution limit of the
simulations is suppressed at early epochs. Thus, when we 
increase the resolution in the \LCDM simulations star formation occurs
in objects which did not form at the lower resolution.  However, in
the power spectrum of the warm dark matter cosmology scales, $\lambda
\approxlt 2.5 \Mpc$, are intrinsically suppressed. Consequently, we see very little
resolution dependence in the star formation history of the \LWDM
simulations as the resolution of the $34^3$ simulations is $\lambda =
0.35 \Mpc$ and therefore well below the cut off in the power spectrum.
If we were to increase our resolution still further we
would expect that the early star formation in the \LCDM simulation
to increase whilst very little change would be expected in
the history of the \LWDM simulations.

If we now compare the effect of resolution on the redshift where half of
the stars are formed, $z^*_{1/2}$, in the two cosmologies, we see that it
increases with resolution in the \LCDM cosmology but remains constant
for the \LWDM cosmology (Table \ref{rescomp}).  Thus if we extrapolate
to higher resolution we conclude that $z^*_{1/2}$ in the \LWDM
cosmology will not change but the discrepancy between the formation
redshifts in the two cosmologies would increase and could possibly
produce a significant observable difference.

\section{Detailed comparison with observations of elliptical galaxies} 
\label{obs}

\begin{figure*}
\begin{center}
\epsfig{file=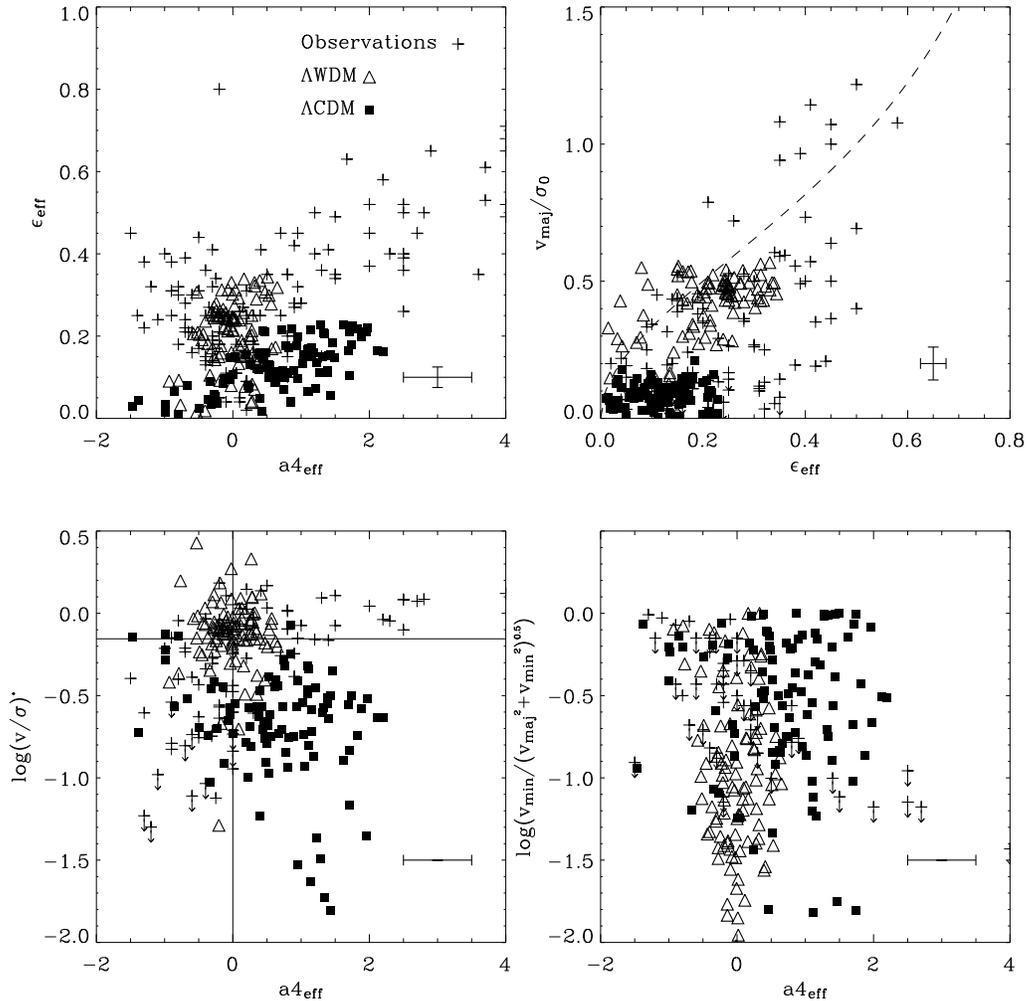, width=0.8\textwidth}
\caption{Kinematical and photometric properties for 100 random projections
of the simulated \LWDM and \LCDM galaxies in the $50^3$ simulation.
Open triangles show the values for the \LWDM galaxy while filled boxes
indicate the values for the \LCDM galaxy. Observed values are
indicated by plus signs, arrows indicate upper limits (Data kindly 
provided by Ralf Bender). The errors were derived by statistical
bootstrapping. {\it Upper left panel}: Effective ellipticity
$\epsilon_{\mathrm{eff}}$ of the galaxies versus effective isophotal shape
$a4_{\mathrm{eff}}$. {\it Upper right panel}: Rotational velocity over central
velocity dispersion ($v_{\mathrm{maj}}/\sigma_0$) versus $\epsilon_{\mathrm{eff}}$ . The
dashed line shows the theoretical curve for an oblate isotropic
rotator. {\it Lower left panel}: Anisotropy parameter $(v/\sigma)^*$ 
versus $a4_{\mathrm{eff}}$. {\it Lower right panel}: Amount of minor axis rotation
versus $a4_{\mathrm{eff}}$with $v_{\mathrm{maj}}$ an $v_{\mathrm{min}}$
being the maximum velocity along the major and minor axis,
respectively.\label{ellobs}} 
\end{center}
\end{figure*}

The analysis of the previous sections has shown that the simulated galaxies are
hot spheroidal stellar systems. In this section we present a
detailed comparison of the two galaxies formed in the $50^3$ \LCDM
and \LWDM simulations with observations of giant elliptical galaxies
with respect to observed characteristic 
photometric and kinematical properties, e.g. surface density profile,
isophotal deviation from perfect ellipses, velocity  dispersion, and
major- and minor-axis rotation. In addition we analyse the
line-of-sight velocity distribution (LOSVD) of both galaxies. The 
methods  used to analyse the simulated galaxies closely follows the
analysis used by observers and is described in detail in
\citet{NB2003}. Only a brief description will be given here. We
created an artificial image of the simulated galaxies by  
binning the central 35 \kpc into $128 \times 128$ pixels 
and smooth it with a Gaussian filter of standard deviation 
1.5 pixels. Using this image we derived the surface brightness
distribution and the best fitting Sersic-profile (\citealp{Ser1968})
$\Sigma = \Sigma_0 \exp({-r^{1/n_{\mathrm{ser}}})}$, where
$n_{\mathrm{ser}}$ is the Sersic-index. For $n_{\mathrm{ser}} = 1$ the profile is
exponential, $n_{\mathrm{ser}} =4$ parameterises the de Vaucouleurs $r^{1/4}$
profile. The isophotes were analysed with respect to the fourth-order
cosine deviation, $a_4$, from perfect ellipses
\citep{1988A&AS...74..385B}. For every projection of a 
simulated galaxy we defined an effective $a_4$-coefficient,
$a4_{\mathrm{eff}}$, as the mean value of $a_4$ between $0.25 r_e$ and
$1.0 r_e$, with $r_e$ being the projected spherical half-light radius.
For the moment we are treating (M/L) as constant for all stellar
particles regardless of when they were created, or the metallicity of
the gas from which they originated.  

The characteristic ellipticity $\epsilon_{\mathrm{eff}}$ for each
projection was defined as the isophotal ellipticity at $1.5 r_e$. The
central velocity dispersion $\sigma_0$ was determined as the average
projected velocity dispersion of all stellar particles inside a
projected galactocentric distance of $0.2 r_e$. The
characteristic rotational velocity along the major and the minor axis
were the projected rotational velocities at $1.5 r_e$ and $0.5 r_e$,
respectively.

We analysed the surface brightness distributions of 100 random
projections of the \LCDM and \LWDM galaxy using the Sersic-profile. 
The \LCDM galaxy has a half-mass radius of $r_{\mathrm{eff}} \approx
3.2\ kpc$ and a mean Sersic-index of $n_{\mathrm{ser}}= 1.8 \pm 0.1$ which
is close to the observed lower limit for intermediate-mass 
ellipticals at the same effective radius \citep{1993MNRAS.265.1013C}.    
The \LWDM galaxy shows a more exponential profile with a mean
Seric-index of $n_{\mathrm{ser}}= 1.5 \pm 0.15$ at a smaller effective
radius of  $r_{\mathrm{eff}} \approx 2.2\ kpc$ again consistent with
lower observational limits (see Fig. \ref{global}). Interestingly,
the observed trend for more exponential galaxies to have smaller
effective radii is followed. Combined with the ellipticities and
isophotal deviations derived in the following section the photometric
properties of both galaxies agree with observed ellipticals (see
\citealp{1993MNRAS.265.1013C}).    

\begin{figure}
\begin{center}
\epsfig{file=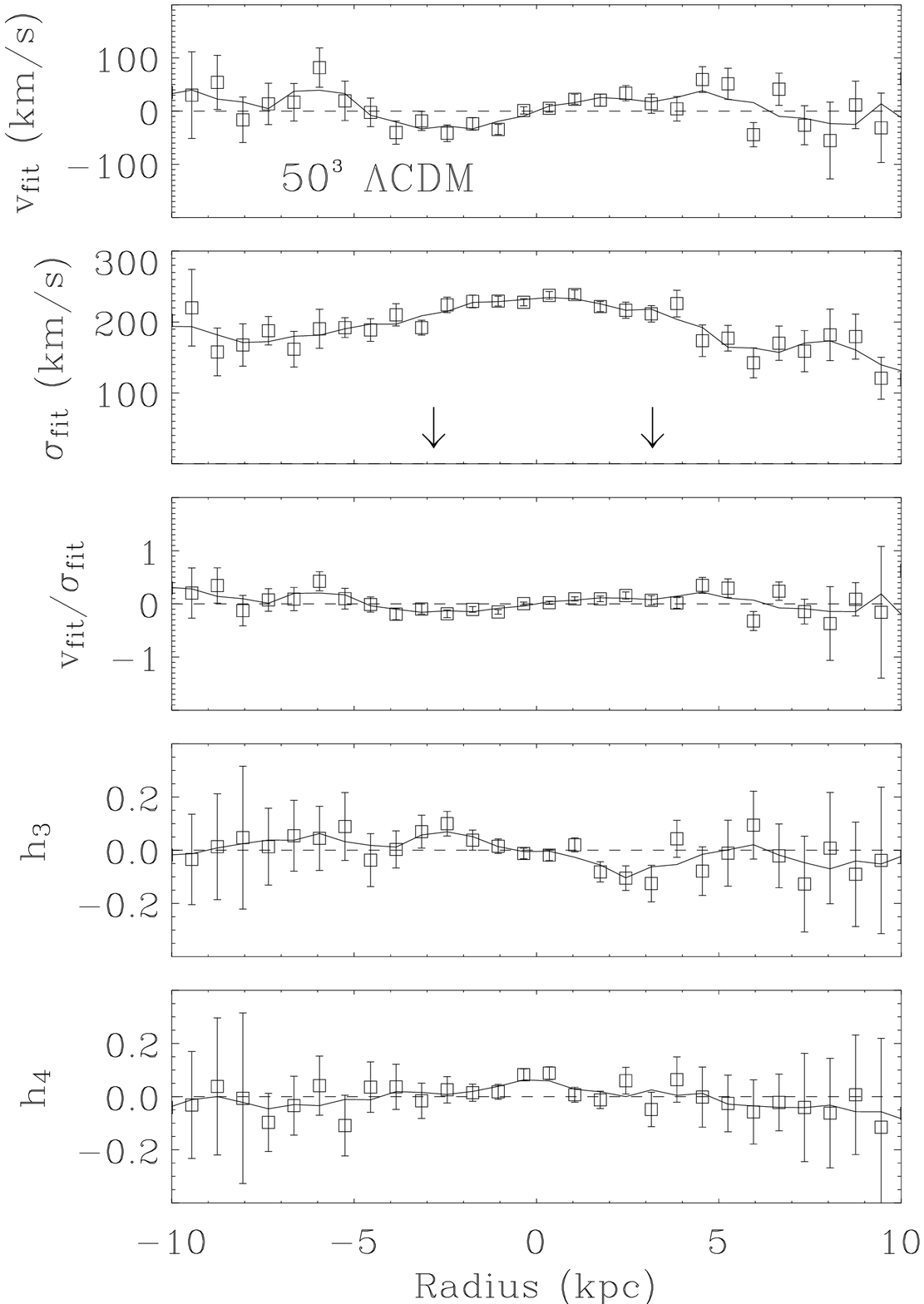, width=0.45\textwidth}
\caption{Analysis of the LOSVD of the $50^3$ \LCDM galaxy as measured along
   a slit aligned with the major axis of the moment of inertia tenor
   of the main stellar body (open squares). The fitted local velocity
   $v_{\mathrm{fit}}$, local velocity dispersion $\sigma_{\mathrm{fit}}$, 
    $v_{\mathrm{rot}}/\sigma_{\mathrm{fit}}$, $h_3$, and $h_4$ are
   plotted versus radius. The lines show the
   smoothed profiles, respectively. The effective radius of the galaxy,
   $r_{\mathrm{eff}} \approx 3.2\ kpc$ is indicated by the arrows. The
   individual error-bars were derived by
   bootstrapping. \label{lcdmlosvd}} 
\end{center}
\end{figure}

\begin{figure}
\begin{center}
\epsfig{file=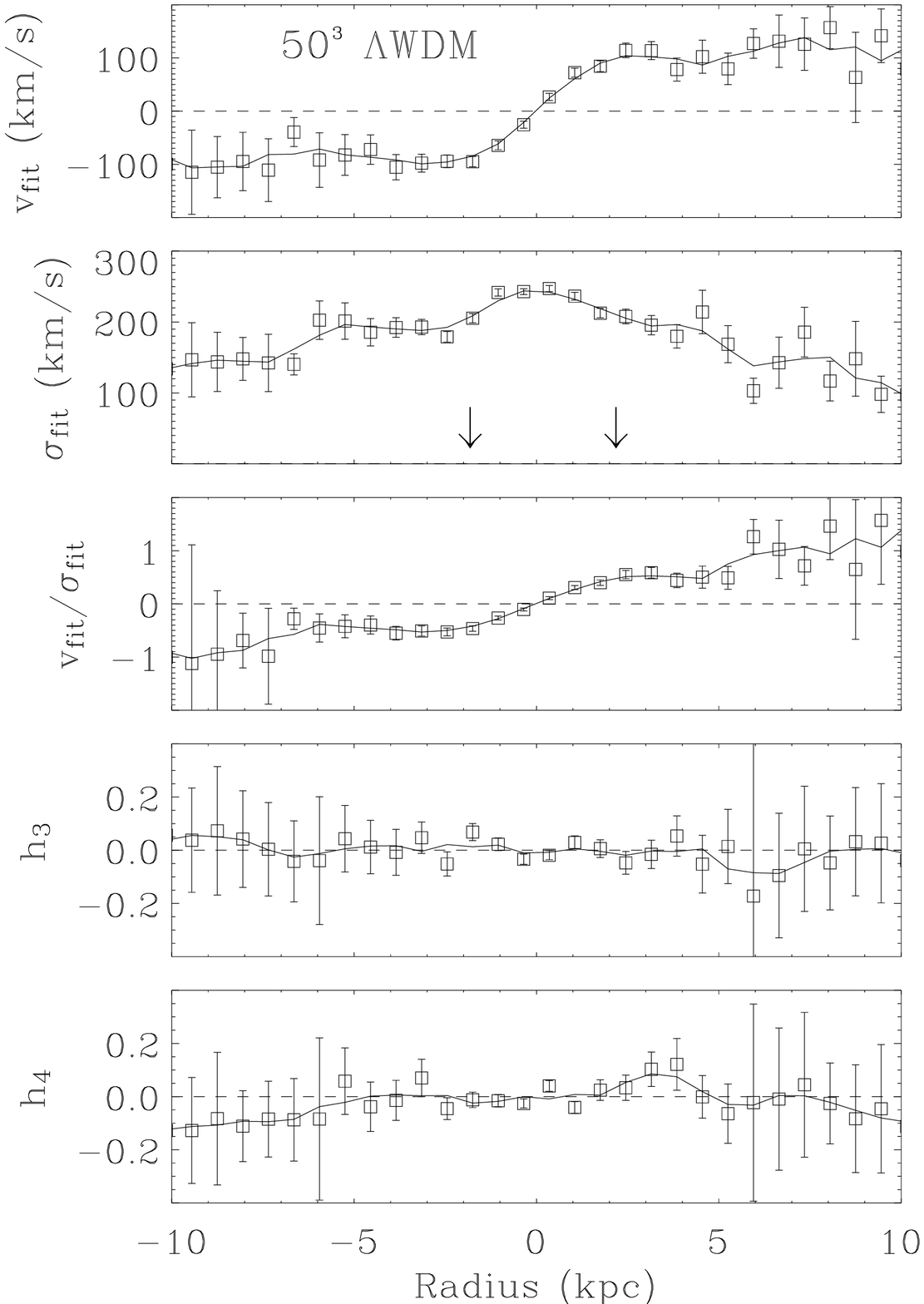, width=0.45\textwidth}
\caption{Same as Fig. \ref{lcdmlosvd} but for the $50^3$ \LWDM
   galaxy. The effective radius of the galaxy,
   $r_{\mathrm{eff}} \approx 2.2\ kpc$ is indicated by the arrows. \label{lwdmlosvd}} 
\end{center}
\end{figure}

In Fig. \ref{ellobs} we show the global analysis of the two galaxies
resulting from the $50^3$ \LCDM (filled boxes) and \LWDM (open
triangles) simulations seen along 100 random lines-of-sight. The
observations of real elliptical galaxies are indicated by plus signs
(data kindly provided by Ralf Bender). In the upper left plot of
Fig. \ref{ellobs}  the projections of the \LCDM galaxy in the
$a4_{\mathrm{eff}}$-$\epsilon_{\mathrm{eff}}$ plane are shown. 
The galaxy looks almost round with a maximum ellipticity of 0.2 and a
peak in the ellipticity distribution around $\epsilon = 0.1$. It has
predominantly disky ($a4_{\mathrm{eff}} > 0$) deviations from elliptical
isophotes and there is a trend for more disky projections to have
higher ellipticities. The \LWDM galaxy is more elongated and the
distribution of ellipticities peaks at $\epsilon = 0.2$ with a maximum
of $\epsilon = 0.35$. The isophotes are either slightly boxy or
disky. Taking the errors of the $a_4$ analysis into account the
isophotal shape scatters around zero. The area 
covered by the projections of both galaxies in the
$a4_{\mathrm{eff}}$-$\epsilon_{\mathrm{eff}}$ plane is consistent with 
the observed distribution. However, the simulations do not reproduce
galaxies with $\epsilon_{\mathrm{eff}} > 0$ and $a4_{\mathrm{eff}} > 2
$ or $a4_{\mathrm{eff}} < -1$.

With respect to rotation, the \LCDM galaxy is a slow rotator with
$(v_{maj}/\sigma_0) < 0.2$ (upper right plot in Fig. \ref{ellobs})
and is likely to be flattened by anisotropic velocity dispersions as
the anisotropy parameter $(v_{maj}/\sigma_0)^*$  is in general well
below 0.7. As this galaxy shows predominantly disky isophotes it
occupies a region in the $(v_{maj}/\sigma_0)^*$-$a4_{\mathrm{eff}}$
plane where no real galaxy is observed (lower left plot in Fig.
\ref{ellobs}). Additionally, in the
$(v_{maj}/\sigma_0)$-$\epsilon_{\mathrm{eff}}$-plane the projections  
cover the area of observed massive boxy ellipticals which are much
more luminous than the galaxies presented here. In contrast, the \LWDM
galaxy has projected $(v_{maj}/\sigma_0)$-values higher than 0.5,
which is consistent with models for rotationally supported galaxies
(dashed line in upper right plot of Fig. \ref{ellobs}). The galaxy
also is fairly isotropic with $(v_{maj}/\sigma_0)^*$ scattering around
zero and agrees well with observed isotropic galaxies with small
deviations from elliptical isophotes. None of the projections of the
simulated galaxies falls in the area of boxy anisotropic or disky
isoptropic ellipticals. Both simulated galaxies show minor-axis
rotation (lower right plot of Fig. \ref{ellobs}). The \LCDM galaxy
however disagrees with observations as it shows disky isophotes at the
same time. Observed disky ellipticals show only weak minor axis
rotation. \\

\begin{figure}
\begin{center}
\epsfig{file=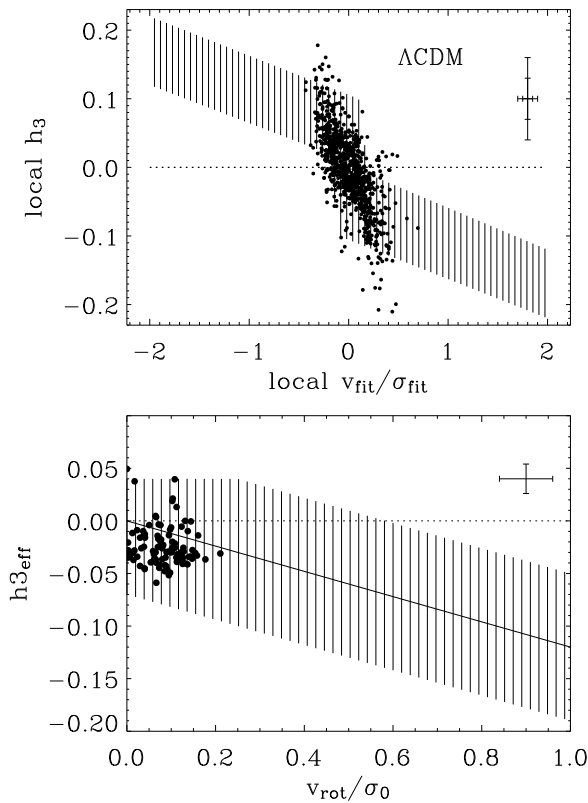, width=0.45\textwidth}
\caption{{\it Upper plot:} Local correlation between $h_3$ and
$v_{\mathrm{fit}}/\sigma_0$ for the \LCDM galaxy. The observed data
points cover the region indicated by the shaded area. The large
error bar indicates the errors for $v_{\mathrm{fit}}/\sigma_0 > 0.2$.
{\it Lower plot:} Global correlation between $h3_{\mathrm{eff}}$ and
$v_{\mathrm{fit}}/\sigma_0$ The observed correlation
(\citealp{BSG1994}) is shown by a straight line. The maximum spread
around this correlation is indicated by the shaded area. The dotted
lines indicate the zero line. \label{h3vs_lcdm}}
\end{center}
\end{figure}
\begin{figure}
\begin{center}
\epsfig{file=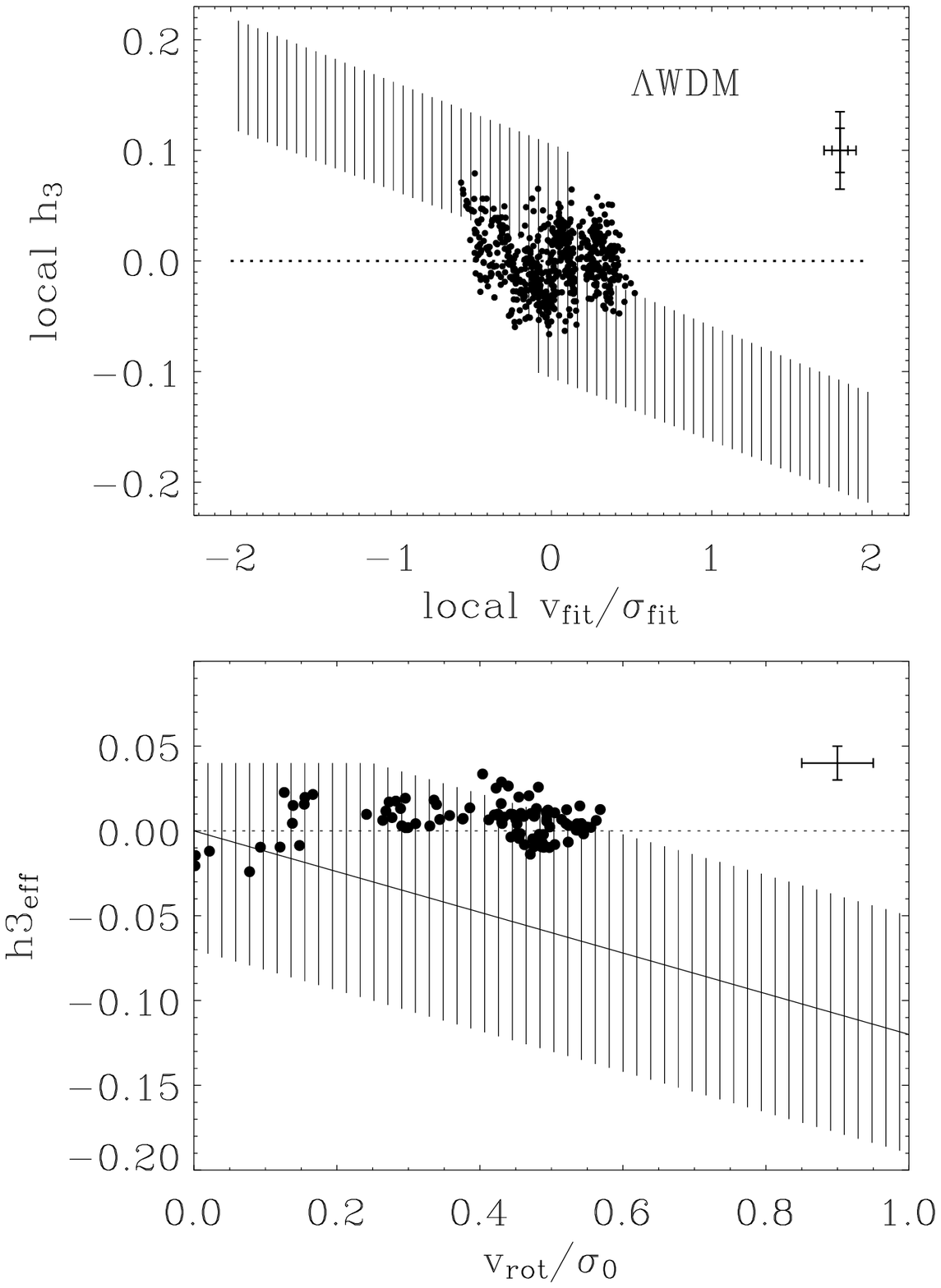, width=0.45\textwidth}
\caption{ Same as Fig. \ref{h3vs_lcdm} but for the \LWDM galaxy.\label{h3vs_lwdm}}
\end{center}
\end{figure}

To analyse the LOSVDs of the simulated galaxies in more
detail we placed a slit with a width of 1.5\kpc and a grid spacing of
0.5\kpc along the apparent long axis of each projected
remnant. Thereafter all particles falling within each grid cell
were binned in velocity along the line-of-sight. The line-of-sight velocity
profiles for each bin along the grid were then parameterised by
a Gaussian plus third- and fourth-order Gauss-Hermite basis 
functions \citep{vdMF1993,BSG1994,BB2000}. Figs. \ref{lcdmlosvd} and
\ref{lwdmlosvd} show the kinematic parameters
($\sigma_{\mathrm{fit}}$, $v_{\mathrm{fit}}$, $h_3$,$h_4$) along the 
long axes of the \LCDM and \LWDM galaxy, respectively. $h_3$ and $h_4$
are the amplitudes of the third- and fourth order  Gauss-Hermite
functions. For $h_3= 0$ and $h_4= 0$ the resulting velocity profile
is a Gaussian. For asymmetric profiles with the prograde (leading)
wing steeper than the retrograde (trailing) one, $h_3$ and
$v_{\mathrm{fit}}$ have opposite signs. When $v_{\mathrm{fit}}$ and
$h_3$ have the same sign, the leading wing is broad and the trailing 
wing is narrow. LOSVDs with $h_4 > 0$ have a 'triangular' or
peaked shape, here the distribution's peak is narrow with broad
wings. Flat-top LOSVDs have $h_4 < 0$, where the peak is broad
and the wings are narrow.

The \LCDM galaxy shows very little rotation inside its effective 
radius (Fig. \ref{lcdmlosvd}). The velocity dispersion inside
$r_{\mathrm{eff}}$ is flat and falls off slowly at larger
radii therefore $v_{\mathrm{fit}}/\sigma_{\mathrm{fit}}$ never exceeds
0.2. The asymmetry of the LOSVD parameterised by $h_3$ is
anticorrelated with $v_{\mathrm{fit}}$ as it is observed for real
ellipticals \citep{BSG1994}. The local correlation between $h_3$ and
$v_{\mathrm{fit}}/\sigma_{\mathrm{fit}}$ inside $e_{\mathrm{eff}}$ for 100
random projections of the simulated galaxy is shown 
in upper plot of Fig. \ref{h3vs_lcdm}.  $h_3$ and
$v_{\mathrm{fit}}/\sigma_{\mathrm{fit}}$ agree well with observed data
(indicated by the shaded area) for
$v_{\mathrm{fit}}/\sigma_{\mathrm{fit}} < 0.2$ and are in general  
anticorrelated.  This is also reflected in a negative 
effective $h_3$ value, $h3_{\mathrm{eff}}$, (lower plot in Fig.
\ref{h3vs_lcdm}) for 
almost all projections, where $h3_{\mathrm{eff}}$ is defined as the
mean value of $h_3$ 
inside one effective radius (see \citealp{BSG1994}). The points follow
the observed correlation between $h3_{\mathrm{eff}}$ and
$v_{\mathrm{rot}}/\sigma_0$.\\ 

The \LWDM galaxy, as we have already seen in Fig. \ref{ellobs},
shows significant rotation 
(Fig. \ref{lwdmlosvd}) with the rotation velocity flattening out at
around 100 km/s beyond the effective radius. The velocity
dispersion profile is more peaked than for the \LCDM case. This results 
in local $v_{\mathrm{fit}}/\sigma_{\mathrm{fit}}$ as high as 0.6. The
parameters $h_3$ and $h_4$ are consistent with zero inside
$r_{\mathrm{eff}}$. Analysed for 100 projections the local value of
$h_3$ does not correlate with 
$v_{\mathrm{fit}}/\sigma_{\mathrm{fit}}$ which is in contradiction to
observations (upper  
plot of Fig. \ref{h3vs_lwdm}). $h3_{\mathrm{eff}}$ (lower plot in Fig.
\ref{h3vs_lwdm}) tends to stay positive and does not follow the
observed trend, however it is broadly consistent with observations.\\    
\begin{figure}
\begin{center}
\epsfig{file=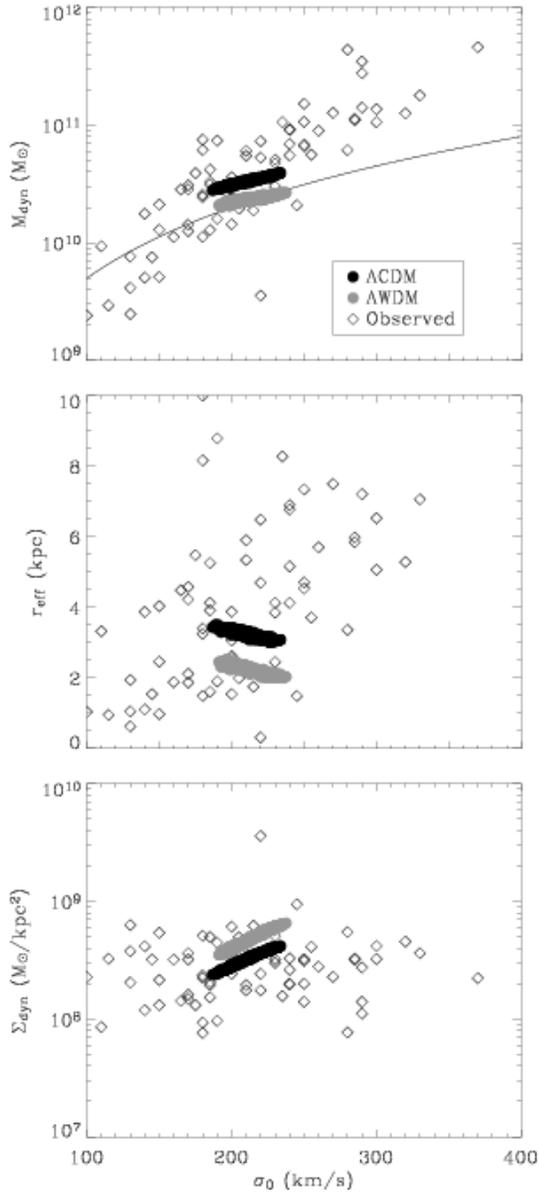, width=0.45\textwidth}
\caption{Dynamical mass $M_{\mathrm{dyn}} =
r_{\mathrm{eff}} \sigma_0^2 / G$, effective radius $r_{\mathrm{eff}}$,
and dynamical surface-brightness, $\Sigma_{\mathrm{dyn}}$, versus
central velocity dispersion $\sigma_0$. Black dots represent
projections of the \LCDM, grey dots of the \LWDM galaxy,
respectively. For the simulated galaxies $r_{\mathrm{eff}}$ is the
projected spherical half-mass radius. \label{fp}}
\end{center}
\end{figure}

\begin{figure}
\begin{center}
\epsfig{file=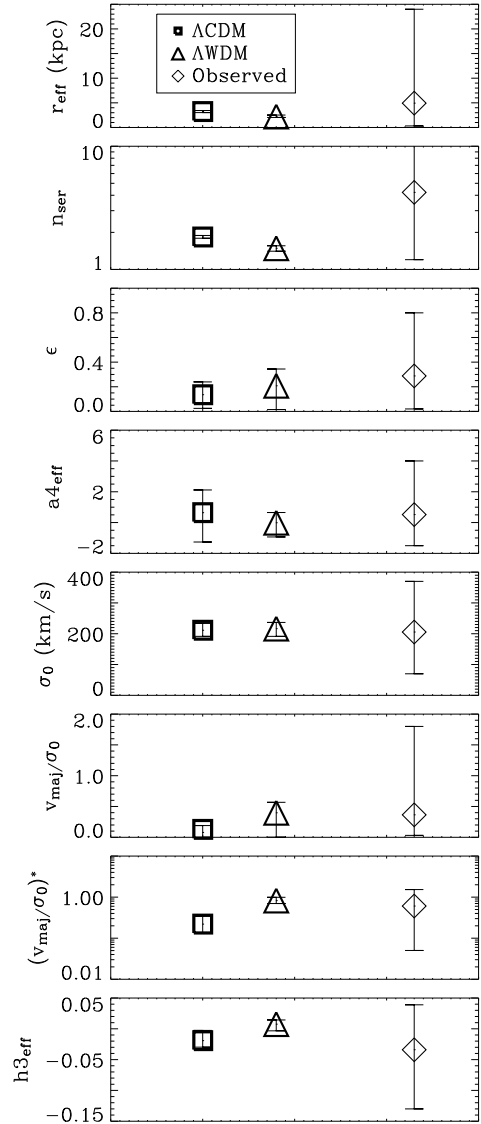, width=0.4\textwidth}
\caption{Mean properties of the simulated \LCDM (squares), \LWDM
(triangles) and observed (diamonds) elliptical galaxies. The maximum
spread in the data is indicated by the error bars. For the two simulated
galaxies the spread is due to projection effects. Observational data
for the effective radius ($r_{\mathrm{eff}}$), ellipticity
($\epsilon$), effective isophotal shape ($a4_{\mathrm{eff}}$), rotation
velocity ($v_{\mathrm{maj}}$) and velocity dispersion ($\sigma_{0}$)
have been kindly provided by Ralf Bender. Data for the Seric index
$n_{\mathrm{ser}}$ have been taken from \citet{1993MNRAS.265.1013C} and
for the LOSVD asymmetries ($h3_{\mathrm{eff}}$) from
\citet{BSG1994}. \label{global}} 
\end{center}
\end{figure}

In summary, the \LWDM galaxy can be interpreted as a modestly rotating
isotropic hot stellar system. Its global photometric and kinematical
properties are consistent with observations of isotropic
ellipticals. The galaxy does, however, not show any strong sign of
the fine-structure that would be typical for rotationally supported
elliptical galaxies. Its isophotal shape does not deviate
significantly from being elliptical and there is no evidence that the
LOSVDs show a correlated deviation from a Gaussian. This may be due,
at least in part, to an insufficient number of particles in the
simulation. \\

The \LCDM galaxy shows signatures for being a combined
disk/bulge system (except for the very low net rotation rate). The
photometric decomposition (Fig. \ref{surf}) 
presents evidence for a disk-like sub-component. This is supported by
the predominantly disky shape of the galaxy. The anticorrelation
between $h_3$ and $v/\sigma$ provides further kinematical evidence for
a combined disk-bulge system \citep{BSG1994, 2001ApJ...555L..91N}.
From the rotation curves, the round shape and the anisotropic
kinematics of the galaxy it becomes clear that the ``disk'' can
contribute only a small fraction to the total mass of the system.
However, the \LCDM galaxy rotates too slowly to be consistent with
observations and there is no sign of a disk in the kinematic
decomposition.  It forms a disky anisotropic system which is not
observed.\\

Due to the uncertain mass-to-light ratios of the simulated galaxies we
use the virial theorem to define a dynamical masses $M_{\mathrm{dyn}}=
r_{\mathrm{eff}} \sigma_0^2 / G$ using the half-mass radius
$r_{\mathrm{eff}}$ and the central velocity dispersion $\sigma_0$ of
every projected galaxy. This mass is used to define a dynamical
surface density, $\Sigma_{\mathrm{dyn}} =
M_{\mathrm{dyn}}/r_{\mathrm{eff}}^2 $, to create a simulated  
Fundamental Plane where the effective surface brightness $SB_e$ is
replaced by $\Sigma_{\mathrm{dyn}}$. In the same way we derive
dynamical masses for observed ellipticals using the observed effective
radius and velocity dispersion. Fig. \ref{fp} shows a comparison of
observed ellipticals with the simulated \LCDM and \LWDM galaxies. Both
galaxies are consistent with observations. The \LWDM galaxy has a
smaller dynamical mass and has a higher effective surface brightness
than the \LCDM galaxy. \\

Figure \ref{global} shows a comparison of the mean global projected
properties of the simulated galaxies to the global properties of
giant elliptical galaxies. The maximum spread in the data and in the
observations is indicated by the error bars. It becomes clear from
this figure that we produced ``ordinary'' ellipticals in our
simulations which compare very well to the average properties of giant
ellipticals. 

The low number of stellar particles within the half-mass radii of the
simulated galaxies ($\approx 7000$) makes it too early to draw any
final conclusion on the real structure of intermediate mass
ellipticals in \LWDM or \LCDM cosmologies. Especially as the \LCDM
simulations are not yet numerically converged.  Relaxation effects in
particular might significantly influence the dynamics of the systems
(see \citealp{astro-ph/0304549}) 

The present resolution is already sufficiently high to perform a
 kinematical and photometric analysis within reasonable error limits 
 (all errors have been determined by applying the statistical
 bootstrapping method (see \citealp{1994ApJ...427..165H}) both for global and local
 measurements (see {\it e.g.} Figs. \ref{ellobs},
 \ref{lcdmlosvd}). Therefore we are able to trace the real structure
 of the simulated galaxies at the present resolution. Increasing the
 resolution, however, will very likely change the detailed internal
 stellar structure of the simulated 
 galaxies whereas the differences will presumably be stronger for \LCDM
(as we have shown in the previous section) than for \LWDM cosmologies.
A preliminary analysis of a higher resolution \LCDM supports this conclusion.\\

\section{Conclusions}\label{concs}
We have presented the results of ten $34^3$ particles and two $50^3$
particles simulations of the formation of individual galaxies in a
warm and cold dark matter cosmology with a cosmological constant.  The
sample of low resolution simulations enabled us to compare the global
properties of the dark matter haloes and galaxies formed in the two
different cosmologies. The two galaxies selected for re-simulation
reside in a low density environment. Consequently the models presented
here are are likely to trace the formation of ordinary intermediate
mass giant elliptical or S0/Sa galaxies in the field, for example, the
Sombrero galaxy (M104). The two high resolution simulations have been
used to investigate the assembly histories of the final galaxies and
to compare their internal properties with those of real early type
galaxies.

As expected from the suppression of small-scale structure in the
initial conditions, the \LWDM cosmology produces fewer low mass dark
matter haloes ($M_{halo} < 10^{10} M_{\odot}$) at the present epoch
compared with the \LCDM cosmology \citep{2001ApJ...556...93B}. This
feature is also reflected in the assembly histories of the two $50^3$
simulations. At almost all redshifts, the \LCDM galaxy experiences
more minor mergers with higher mass-ratios (up to 10:1) than the
corresponding \LWDM galaxy. Low accretion rates, and the absence of
any major merger event, are expected since the initial conditions were
selected from low density environments.  In addition to differences in
the  frequency of mergers,  the internal composition of the merging
satellites differs in the two cosmologies. The \LWDM satellites are
always more gas rich. At redshifts $z < 0.3$  the star-to-gas ratio is
about an order of magnitude higher in the \LCDM than in the \LWDM
cosmology. As a result,  below a redshift of two (when both galaxies
have already assembled   $\approx 50\% $ of their final stellar mass)
about 30\% of all present day stars in the \LCDM galaxy have been
accreted by mergers and $\approx 20\%$ of the stars have formed inside
the galaxy. Over the same  period the \LWDM has accreted only $\approx
10\%$ of its present day stars and $\approx 40\%$ have formed within
the galaxy.

A further difference between the two cosmologies is  evident in the
angular momentum evolution of the galaxies.  The halo of the \LCDM
galaxy has higher  angular momentum than the \LWDM halo but they both
show a  similar temporal evolution. In contrast, the average angular
momentum of the stars of the \LCDM galaxies does not change
significantly after a redshift of $z=4$ whereas the stars of the \LWDM
galaxy gain further angular momentum until $z=1$,  with little  evolution
thereafter. Between a redshift of $z=4$ and $z=1$ the \LWDM galaxy is
more gas rich and forms more stars within the galaxy than its \LCDM
counterpart, resulting in a second peak in the star formation rate at $z
\approx 1.5$.

We conclude that the removal of small scale power reduces the
frequency of mergers.  The reduction in the number of massive sub-halo
mergers in the $50^3$ warm dark matter simulation produces a galaxy
with significantly higher angular momentum at z=0. The increase in the 
specific angular momentum of objects in the warm dark matter
simulations suppresses the collapse of gas within the simulations,
which in turn reduces the star formation rate of the central galaxy at
early epochs. 

We performed a photometric and kinematical decomposition of the main
stellar systems in the  $50^3$ simulations. Both galaxies are
dominated by a hot spheroidal component. The global projected
properties of the two $50^3$ \LCDM and \LWDM galaxies resemble those
of real elliptical galaxies. The total masses, density profiles,
effective radii, ellipticities, global values of isophotal shapes and
LOSVD asymmetries are consistent with the average global properties of
giant elliptical galaxies of intermediate masses. This is in contrast
to properties of very massive, anisotropic, and boxy ellipticals with
large effective radii, typically found in clusters of galaxies. As the
physical properties of massive ellipticals differ from those of
ordinary intermediate mass ellipticals with respect to inner
density-profiles, sizes, isophotal shapes and X-ray and radio
properties \citep{1989A&A...217...35B, 1993MNRAS.265.1013C,
1997AJ....114.1771F}, we can conclude that they must have formed in a
denser environment and, therefore had different formation histories.

Investigating the \LWDM galaxy in detail revealed that it appears to
be an isotropic fast rotator with only weak fine structure both in its
isophotal shape (the isophotes are elliptical) and in its LOSVDs (which
on average have a Gaussian shape). The ratio of
$T_{\mathrm{rot}}/T_{\mathrm{rand}}$ is a factor of ten higher for the
\LWDM than for the \LCDM system. Observed rotationally supported 
ellipticals with similar rotational support do, however, show on
average stronger asymmetries in their LOSVDs. The detailed photometric
and global kinematical properties do agree very well with
observations. The half-mass radius of stellar population is 2.3
$kpc$. In contrast, the \LCDM galaxy shows only weak rotation and
appears to have anisotropic velocity dispersions. The isophotes are
predominantly disky and the LOSVDs show the observed trend which
indicates the presence of a weak disk component embedded in the
spheroidal body of the galaxy (\citealp{1990ApJ...362...52R, BSG1994,
2001ApJ...555L..91N}).  However, anisotropic disky systems, like the
\LCDM elliptical, are in general not observed. The half-mass radius of
the \LCDM galaxy is 3.2 $kpc$.  The mean age of the stellar population
of both galaxies is about 10 $Gyrs$ which is in good agreement with
the ages of early type galaxies.

\citet{2003ApJ...590..619M} have recently simulated the formation of
an individual elliptical galaxy in a \LCDM cosmology which is only a little
more massive (by a factor of about 1.5)  than our \LCDM galaxy. However,
their galaxy does not match the global properties of observed giant
ellipticals as its stellar distribution is far too dense. Numerical
resolution is unlikely to cause the difference as their initial
particle masses and softening lengths are comparable to the values
used for the simulations presented here. Their final galaxy, however, has
significantly more stellar particles ($\approx 65000$) than our
galaxies ($\approx 13000$). This is due to differences in the star 
formation algorithms. \citet{2003ApJ...590..619M} split their gas
particles into several stars which can have different masses. As a result
their mean stellar particle mass is a factor of four smaller than our
fixed stellar mass. It is not clear in how far this difference influences the
results,  e.g. due to mass segregation effects in the stellar  
distribution. A further difference is that \citet{2003ApJ...590..619M}
make their stars at a rate proportional to the gas density with a
relatively low efficiency. In the simulations presented here a
gas particle is turned into a star particle as soon as its density is
above a certain threshold for more than a dynamical time. As a result
the gas in the \citet{2003ApJ...590..619M} simulation can collapse to
much higher densities before stars are formed.

In addition we omitted feedback processes in our
simulations. \citet{2003ApJ...590..619M} implemented thermal and
kinematical feedback. Although one might naively expect that the
inclusion of feedback would result in less compact objects, this goes
in the opposite direction to explain the differences between their
results and ours. Another possible explanation for the differences is
the merger history. Our galaxies experience only minor mergers with
mass ratios up to 10:1. The \citet{2003ApJ...590..619M} galaxy
undergoes a late major merger with a mass-ratio of 3:1 which could
effectively drive gas into the centre and convert it into
stars. However, the star formation history is not significantly
different from our \LCDM galaxy which undergoes no major
merger. The merger is therefore  more likely to influence the
dynamics of the system rather than have a major effect on the
concentration of the bulk of the stellar population.

Based on the simulations at two different resolutions in a \LCDM and
\LWDM cosmology it becomes clear that this investigation is not yet
definitive. As the simulations are not yet fully numerically resolved
(especially the \LCDM simulation) we cannot say with confidence whether \LCDM or
\LWDM produces galaxies that most closely resemble real elliptical galaxies. 
In conclusion, we can state that with the simple physics included in
our simulations, it is possible to produce ellipticals in both
cosmologies with global properties that are in good agreement with
observations of intermediate mass giant ellipticals or S0s.  However,
the combination of the detailed properties of our simulated galaxies
(which are very likely to be resolution dependent), like the shape of
the LOSVD or the isophotal shape, differ slightly from observations of
real ellipticals for one combination or the other. Future simulations
at higher resolution will hopefully enable us to determine which, if
either, of the two cosmologies produce galaxies which match the
observations.

\section{Acknowledgements}

Thorsten Naab acknowledges the award of a PPARC PDRA.

\bibliographystyle{mn2e}
\bibliography{refs,thorsten}

\end{document}